\documentstyle[12pt]{article}
\def\beq{\begin{equation}}
\def\eeq{\end{equation}}
\def\bea{\begin{eqnarray}}
\def\eea{\end{eqnarray}}

\def\ba{\begin{array}}
\def\ea{\end{array}}
\def\bce{\begin{center}}
\def\ece{\end{center}}

\def\nonu{\nonumber}

\def\pa{\partial}

\def\ep{\epsilon}

\begin{document}
\begin{titlepage}
\rightline{SNUTP-97-153}
\rightline{UM-TG-204}
\rightline{hep-th/9712005}
\def\today{\ifcase\month\or
January\or February\or March\or April\or May\or June\or
July\or August\or September\or October\or November\or December\fi,
\number\year}
\vskip 1cm
\centerline{\Large \bf M Theory Fivebrane and }
\centerline{\Large \bf Confining Phase of $N=1$ $SO(N_c)$ Gauge Theories }
\vskip 1cm
\centerline{\sc Changhyun Ahn$^{a,}$\footnote{ chahn@spin.snu.ac.kr},  
Kyungho Oh$^{b,}$\footnote{ oh@arch.umsl.edu}
and Radu Tatar$^{c,}$\footnote{tatar@phyvax.ir.miami.edu}}
\vskip 1cm
\centerline{{\it $^a$ Dept. of Physics, Seoul National University,
Seoul 151-742, Korea}}
\centerline{{ \it $^b$ Dept. of Mathematics, University of Missouri-St. Louis,
St. Louis, Missouri 63121, USA}}
\centerline{{\it $^c$ Dept. of Physics, University of Miami,
Coral Gables, Florida 33146, USA}}
\vskip 2cm
\centerline{\sc Abstract}
\vskip 0.2in

The moduli space of vacua for the confining phase of $N=1$ $SO(N_c)$ supersymmetric
gauge theories in four dimensions is analyzed by studying the M theory fivebrane.
The type IIA brane configuration consists of a single NS5 brane, { \it multiple copies } of
NS'5 branes, D4 branes between them, and D6 branes intersecting D4 branes.
We construct M theory fivebrane configuration corresponding to 
the superpotential perturbation where the power of adjoint field is connected to the number of
NS'5 branes. At a singular point in the moduli space where
mutually local dyons become massless, the quadratic degeneracy of the $N=2$ $SO(N_c)$ 
hyperelliptic curve determines whether this singular point gives a $N=1$ vacua or not.
The comparison of the meson vevs in M theory fivebrane configuration
with field theory results turns out
that the effective superpotential by the integrating in method with enhanced gauge
group $SU(2)$ is exact.
 
\vskip 1in
\leftline{Nov., 1997}
\end{titlepage}
\newpage
\setcounter{equation}{0}

\section{Introduction}
\setcounter{equation}{0}

In the last years we have seen how string/M theory can be used to study 
non-perturbative dynamics of  low energy  supersymmetric gauge theories in
various dimensions. 
One of the main motivations is to understand the D(irichet)
brane dynamics where the gauge theory is realized on the worldvolume of D branes.

This work was pioneered by Hanany and Witten \cite{hw} where 
the mirror symmetry of $N=4$ gauge theory in 3 dimensions was described
by changing the location of the Neveu-Schwarz(NS)5 brane in spacetime ( See,
for example,  \cite{bo1,bo2}  ).
As one changes the relative orientation of the two NS5 branes \cite{bar} while keeping
their common 4 spacetime dimensions intact, the $N=2$ supersymmetry is
broken to $N=1$ \cite{egk,egkrs}. 
Using this configuration they \cite{egk} described and checked a stringy
derivation of Seiberg's duality for $N=1$ supersymmetric gauge theory with
$SU(N_c)$ gauge group with $N_f$ flavors in the fundamental representation
which was conjectured some time ago in \cite{se1}.
This result was generalized to brane configurations with orientifolds which 
give $N=1$ supersymmetric theories with gauge group $SO(N_c)$ or 
$Sp(N_c)$ \cite{eva,egkrs} ( See also \cite{bh,bsty,t} for this approach and
\cite{ov,ao,a,ar,aot} for an equivalent geometrical approach ).

The branes in type IIA/IIB string theory were considered to be rigid without any bendings.
As the branes are intersecting each other, a singularity occurs. 
In order to avoid that kind of singularities, a nice
explanation was found by reinterpreting  brane configuration in string theory
from the point of view of M theory by Witten in \cite{w1}. 
Then both the D4 branes
and NS5 branes used in type IIA string theory 
originate from the fivebrane of M theory ( the former is an M theory
fivebrane wrapped over $\bf{S^1}$ and the latter is the one on
$\bf{R^{10} \times S^1}$ ). That is, D4 brane's worldvolume projects to a
 five dimensional  manifold in
$\bf{R^{10}}$ and NS5 brane's worldvolume is located at a point in $\bf{S^1}$ and
fills a six dimensional manifold in $\bf{R^{10}}$. In order to insert  D6 branes 
one has to use a multiple
Taub-NUT space \cite{town}
whose metric is complete and smooth. Therefore,
the singularities 
are removed in 11 dimensions where the picture becomes smooth, the
D4 branes and NS5 branes become the unique M theory  fivebrane and 
the D6 branes are the
Kaluza-Klein monopoles.
The property of $N=2$ supersymmetry 
in four dimensions requires that the worldvolume of M theory fivebrane is 
${\bf{R^{1,3}}} \times \Sigma$ where $\Sigma$ is uniquely identified with the
curves \cite{sw}-\cite{aps1} that 
appear in the solutions to Coulomb branch of the field theory.
Further generalizations of this configuration with orientifolds were 
studied in \cite{lll,bsty1}.
The original work \cite{w1} was appropriated for  understanding  the moduli space for $N=2$
supersymmetric gauge theories. 
In \cite{w2,hoo} ( See also \cite{biksy,ss,dbo} ), this was seen in 
M theory, by considering the possible deformation of the curve $\Sigma$. 

The exact low energy description of $N=2$ supersymmetric $SU(N_c)$ gauge
theories with $N_f$ flavors in 4 dimensions have been found in \cite{hoo}.
They obtained the information regarding the
Affleck-Dine-Seiberg superpotential \cite{ads} for $N_f < N_c$, in M theory approach. 
It has $N_{c}$ branches
corresponding to $N_{c}$ D4 branes and there exist two asymptotic
regions corresponding to two NS5 branes. The M theory  fivebrane \cite{w2} is 
described by the curve
\begin{equation}
t^{2} - 2 C_{N_c} (v, u_{k}) t + \Lambda_{N=2}^{2N_{c} - N_{f}} \prod_{i=1}^{N_f} ( v+m_i ) = 0
\end{equation}
where $v = x^4 + i x^5, t = \mbox{exp} ( -(x^{6} + ix^{10})/R)$ where $x^{10}$
is the eleventh coordinate of M theory compactified on a circle of radius $R$,
$C_{N_c}(v, u_k)$ is a degree $N_c$ polynomial in $v$ with  coefficients
depending on the moduli $u_{k}$ and $m_i ( i=1, 2, \cdots, N_f )$ is the
mass of quark.
By rotating the $N=2$ configuration ( which implies to add
a mass term $\mu_{2}\mbox{Tr}(\Phi^{2})$ where $\Phi$ is the adjoint field ) an
$N=1$ configuration is obtained.    
The asymptotic conditions are changed and the M theory fivebrane is described 
now in an $( v, t, w)$ space where $w = x^{8} + i x^{9}$. 

This approach has been developed further and
used to study the moduli space of vacua of confining phase of $N=1$ supersymmetric
gauge theories in four dimensions \cite{dbo}. In terms of brane configuration of IIA string
theory, this was achieved by taking 
multiples of NS'5 branes rather than a single NS'5 brane. In field theory, this
is done by generalizing to the case of the superpotential
$\Delta W = \sum_{k=2}^{N_{c}} \mu_{k} \mbox{Tr} (\Phi^{k})$. This
perturbation lifts the non singular locus of the $N=2$ Coulomb branch while
at singular locus there exist massless monopoles that can condense due to the perturbation.

In the present work we extend the results of \cite{dbo} to $N=1$ supersymmetric theories
with gauge groups $SO(2N_{c})$ and $SO(2N_{c}+1)$ and also generalize our previous
work \cite{aotsept} which dealt with a single NS'5 brane  
in the sense that we are considering {\it multiple copies} of NS'5 branes. 
We will describe how the field theory analysis obtained in the low energy superpotential
gives rise to the geometrical structure in $(v, t, w)$ space.
For more than one massless dyon\footnote{
What we mean by dyon is a state charged electrically or magnetically or both.}, 
a mismatch is 
found between field theory results which have been studied in \cite{ty,kty}
and brane configuration results for the exact result $W_{\Delta} = 0$.
As in $SU(N_{c})$ case, this implies that the minimal form for the effective
superpotential obtained by ``integrating in" is not exact \cite{intril},
in general, for several massless dyons.

This paper is organized as follows.
In section 2, we summarize some results concerning the 
$N=2$ moduli space of vacua  for 
$SO(N_c)$ supersymmetric gauge theory. By adding tree level superpotential perturbation
$\Delta W$ to $N=2$ superpotential, we can analyze the $N=1$ field theory.
In section 3, we describe the M theory fivebrane configuration corresponding to $N=1$
theory with superpotential perturbation $\Delta W$.
In section 4, we calculate meson vacuum expectation values ( vevs ) 
and the result 
is in complete agreement with
field theory results discussed in section 2, for one massless dyon.
Finally, in section 5, we come to
the conclusions and the outlook in the future directions.

Note that
this techniques of intersecting branes in string/M theory
have been used to obtain much information about 
supersymmetric gauge theories with different gauge groups and in various 
dimensions \cite{ah,ba,k,cvj1,hov,hz,mmm,fs,gomez,cvj2,hk,hsz,nos,hy,noyy,mi,cs,ll,es,aotaug,lpt,gp,bisty,ooguri,alish,ahk,fs1,dhoo}. 

\section{Field Theory Analysis}
\setcounter{equation}{0}

Let us first review some field theory results already obtained in
\cite{aps,as,hanany,hms,ty,kty}.
We claim no originality for most of results presented in this section 
except the
description of the singular point in the moduli space where 
mutually local dyons become
massless and the computation of the generating function for the parameter 
$\mu_{2k}$ ( See, for example, 
(\ref{mu2k}) and (\ref{muodd}) ).

\subsection{$N=2$ Theory}

Let us consider $N=2$ supersymmetric  $SO(N_{c})$ gauge theory
with matter in the ${\bf N_c}$ 
dimensional representation of $SO(N_c)$. In terms
of $N=1$ superfields, $N=2$ vector multiplet consists of a field strength
chiral multiplet $W_{\alpha}^{ab}$ and a scalar chiral multiplet 
$\Phi_{ab}$, both in the
adjoint representation of the gauge group $SO(N_c)$. The quark hypermultiplets
are made of a chiral multiplet $Q^{i}_{a}$ which couples to the 
Yang-Mills fields where 
$i = 1,\cdots ,2N_{f}$ are flavor indices
and $a = 1, \cdots , N_{c}$ are color indices. 
The
$N=2$ superpotential takes the form,
\begin{equation}
\label{super}
W = \sqrt{2} Q^{i}_{a} \Phi_{ab}  Q^{j}_{b} J_{ij} 
+ \sqrt{2} m_{ij} Q^{i}_{a} Q^{j}_{a},
\end{equation}
where $J_{ij}$ is the symplectic metric 
$( {0 \atop -1 }{ 1 \atop 0}  ) \otimes {\bf 1_{N_f \times N_f}} $
used to raise and lower 
$SO(N_{c})$ flavor indices ( $
{\bf 1_{N_f \times N_f}}$ is the $N_f \times N_f$ identity matrix )
and $m_{ij}$ is a quark mass matrix
$\label{mass}( { 0 \atop 1 }{  1 \atop 0 }  ) 
\otimes \mbox{diag} ( m_{1}, \cdots, m_{N_f} ) $.
Classically, the global symmetries are the flavor
symmetry $Sp(2N_{f})$ when there are no quark masses, in addition to  
$U(1)_{R}\times SU(2)_{R}$ chiral R-symmetry. 
The theory is asymptotically free for the region $N_{f} < N_{c}-2$ and generates
dynamically a strong coupling scale $\Lambda_{N=2}$ where we denote the $N=2$ 
theory
by indicating it in the subscript of $\Lambda$.
The instanton factor
is proportional to $\Lambda_{N=2}^{2N_{c}-4-2N_{f}}$. Then the 
$U(1)_{R}$ symmetry is anomalous and is broken down to a discrete 
$\bf{Z_{2N_{c}-4-2N_{f}}}$ symmetry by instantons.

The $[N_c/2]$ complex dimensional moduli space of vacua 
contains the Coulomb and Higgs branches\footnote{ We denote $[N_c/2]$ by
the value of integer part of $N_c/2$.}.
The Coulomb branch is parameterized by the gauge invariant order parameters
\bea
u_{2k}=<\mbox{Tr}(\phi^{2k})>, \;\;\;\;\; k=1, \cdots, 
[N_c/2],
\label{u2k}
\eea 
where $\phi$ is the scalar field in $N=2$ chiral multiplet.
Up to a gauge transformation $\phi$  can be skew diagonalized to a
complex matrix, 
$<\phi>=\mbox{diag} ( A_1, \cdots, A_{[N_c/2]} )$ where $A_i=
( { 0 \atop  i a_i }{  -i a_i \atop  0}  )$.
At a generic point the vevs of 
$\phi$ breaks the $SO(N_c)$ gauge symmetry
to $U(1)^{[N_c/2]}$ and the dynamics of the theory is that of an 
Abelian Coulomb phase. The Wilsonian effective Lagrangian in the low
energy can be made of the multiplets of $A_i$ and $W_i$ where
$i=1, 2, \cdots, [N_c/2]$. 
If $k$ $a_i$'s are equal and nonzero then there 
exists an enhanced $SU(k)$ gauge symmetry. On the other hand  when they are also zero, 
there is an enhanced $SO(2k)$ or $SO(2k+1)$ depending on whether 
$N_c$ is even or odd.
The property of $N=2$ supersymmetry implies that there are no perturbative corrections
beyond one loop and there exist nonperturbative instanton corrections.

The quantum moduli space is described by a family of genus $2N_c-1$
hyperelliptic spectral curves\footnote{ We will use the subscript  
``even" for the quantity 
corresponding to $SO(2N_c)$, ``odd" for $SO(2N_c+1)$. 
Otherwise they have common
expression.} \cite{as,hanany,dkp}
with associated 
meromorphic one forms,
\bea 
y_{even}^2 & = & C^2_{2N_c}(v^2)-\Lambda_{N=2}^{4N_c-4-2N_f} v^4 
\prod_{i=1}^{N_f}(v^2-m_i^2) \;\;\;\mbox{for} \;\;SO(2N_c), \nonu  \\
y_{odd}^2 & = & C^2_{2N_c}(v^2)-\Lambda_{N=2}^{4N_c-2-2N_f} v^2
\prod_{i=1}^{N_f}(v^2-m_i^2) \;\;\; \mbox{for} \;\;SO(2N_c+1)
\label{curve}
\eea
where $C_{2N_c}(v^2)$ is a degree $2N_c$ polynomial in $v$ with  coefficients
depending on the moduli $u_{2k}$ appearing in (\ref{u2k})  and 
$m_i ( i=1, 2, \cdots, N_f )$ is the
mass of quark, that is, nonzero quark mass matrix element.
Note that the polynomial $C_{2N_c}(v^2)$ is an even function of $v$ which will be identified 
with a complex coordinate $( x^4, x^5 ) $ directions in spacetime in next section and is given by 
\bea
C_{2N_c}(v^2)= v^{2N_c} +\sum_{i=1}^{N_c} s_{2i} v^{2(N_c-i)}=
\prod_{i=1}^{N_c} (v^2-a_i^2),
\label{poly}
\eea
where $s_{2k}$ and $u_{2k}$ are related each other  by so-called Newton's formula
\bea
2k s_{2k}+ \sum_{i=1}^{k} s_{2k-2i} u_{2i} =0, \;\;\; k=1, 2, \cdots, N_c
\label{newton}
\eea
with $s_0=1$. Recall that the symmetric polynomial $s_{2k}$ in $a_i^2$ is
\bea
s_{2k}= (-1)^k \sum_{i_1 < \cdots < i_k} a_{i_1}^2 \cdots
a_{i_k}^2
\eea
at the classical level. From this recurrence relation,
we obtain
\bea
\frac{\pa s_{2j}}{\pa u_{2k}}=-\frac{1}{2k} s_{2(j-k)}
 \;\;\; \mbox{for} \;\;\; j \geq k
\label{su}
\eea
which will be used later.
When $2r$ branch points of (\ref{curve}) coincide, the Riemann surface 
degenerates as we
vary the moduli, giving a singularity in the effective action and 
there exists an unbroken $SO(2r)$
or $SO(2r+1)$ enhanced gauge symmetry. On the submanifold with all 
but $(N_c-r)/2$ of the $a_i$ being
zero ( when $N_c-r$ is even ), the curve becomes
\bea
y_{even}^2 & = & v^{4r} \left( C^2_{2(N_c-r)}(v^2)-\Lambda_{N=2}^{4N_c-4-2N_f}
v^{2N_f-4r+4} \right),  \\
y_{odd}^2 & = & v^{4r+2} \left( C^2_{2N_c-2r-1}(v^2)-
\Lambda_{N=2}^{4N_c-2-2N_f}
v^{2N_f-4r} \right)
\eea
for massless matter. By absorbing the factor $v^{4r}( v^{4r+2} )$
into the new variable ${\tilde{y}}_{even}( {\tilde{y}}_{odd} )$
we will study the property of singular point in the moduli space.

\subsection{ Breaking $N=2$ to $N=1$}

We are interested in a microscopic $N=1$ theory mainly in a phase
with a single confined photon
coupled to the light dyon hypermultiplet while the photons for the rest are free.
By taking a tree level superpotential perturbation  
$\Delta W$ of \cite{kty} made out of the Casimirs of the adjoint fields in the vector
multiplets to the $N=2$ 
superpotential (\ref{super}), the $N=2$ supersymmetry 
can be broken to $N=1$ supersymmetry.  That is,
\bea
W = \sqrt{2} Q^{i}_{a} \Phi_{ab}  Q^{j}_{b} J_{ij} 
+ \sqrt{2} m_{ij} Q^{i}_{a} Q^{j}_{a}+\Delta W
\label{superpotential}
\eea
where\footnote{ Our $\mu_{2k}$ is the same as their $g_{2k}/2k$ in \cite{kty}. } 
$\Phi$ is the adjoint $N=1$ superfields in the $N=2$ vector multiplet and
\bea
\Delta W_{even} & = & \sum_{k=1}^{N_c-2} \mu_{2k} \mbox{Tr} (\Phi^{2k}) +
\mu_{2(N_c-1)} s_{2(N_c-1)} +
\lambda \mbox{Pf} \Phi, \nonu \\
\Delta W_{odd} & = & \sum_{k=1}^{N_c-1} \mu_{2k} \mbox{Tr} (\Phi^{2k})+
\mu_{2N_c} s_{2N_c}. 
\label{delw}
\eea
Here there exists an extra invariant quantity $\mbox{Pf} \Phi=
\frac{1}{2^{N_c} N_c !} \ep_{i_1 j_1 \cdots i_{N_c} j_{N_c} }
\Phi^{i_1 j_1} \cdots \Phi^{i_{N_c} j_{N_c}}$ 
when $N_c$ is even while $\mbox{Pf} \Phi$ vanishes for odd $N_c$.
Note that the $\mu_{2(N_c-1)}$ term is not associated with $u_{2(N_c-1)}$ but
$s_{2(N_c-1)}$ which is proportional to the sum of $u_{2(N_c-1)}$ and the polynomials
of other $u_{2k} ( k < N_c-1 )$ according to (\ref{newton}) ( Similar argument for odd $N_c$ ).
Then microscopic $N=1$ $SO(N_c)$ gauge theory is obtained from $N=2$ $SO(N_c)$
Yang-Mills theory perturbed by $\Delta W$.

$\bullet$ Pure Yang-Mills Theory

Let us first study $N=1$ pure $SO(N_c)$ Yang-Mills theory with tree level 
superpotential (\ref{delw}).
Near the singular points where monopole
singlets charged under $U(1)$ factors become massless,
the macroscopic superpotential of the theory is given by
\bea
W_{even}= \sqrt{2} \sum_{i=1}^{N_c-1} M_i A_i M_i + \sum_{k=1}^{N_c-2}
\mu_{2k} U_{2k} +\mu_{2(N_c-1)} S_{2(N_c-1)} + \lambda U
\label{supereven}
\eea
and 
\bea
W_{odd}= \sqrt{2} \sum_{i=1}^{N_c-1} M_i A_i M_i + \sum_{k=1}^{N_c-1}
\mu_{2k} U_{2k}+
\mu_{2N_c} S_{2N_c}.
\eea
We denote by $A_i$ the $N=2$ chiral superfield of $(N_c-1)$ $N=2$ $U(1)$
gauge multiplets, by $M_i$ those of $N=2$ dyon hypermultiplets, 
by $U_{2k}$ the chiral superfields corresponding to $\mbox{Tr} (\Phi^{2k}) $,
by $S_{2k}$ the chiral superfields which are related to $U_{2k}$ through (\ref{newton}),
and by $U$ the one corresponding to $\mbox{Pf} \Phi$, 
in the low energy theory. The vevs of the lowest
components of $A_i, M_i, U_{2k}, S_{2k}, U$ are written as $a_i, m_{i, dy}, u_{2k}, s_{2k}, u$
respectively.
Recall that $N=2$ configuration is invariant under the group $U(1)_R$
and $SU(2)_R$ corresponding to the chiral R-symmetry of the field theory we mentioned
last subsection. However,
in $N=1$ theory $SU(2)_R$ is broken to $U(1)_J$.
In order to the theory to be consistent
we should specify the charges of $U(1)_R \times U(1)_J$ of the 
fields and parameters as follows.
\bea
\begin{array}{cccl}
&U(1)_R&U(1)_J&\\
A_i & 2 & 0 & \\
M_i M_i & 0 & 2 & \\
\mu_{2k} & 4- 4k & 4 \\
U_{2k} & -2+4k & -2 \\
S_{2k} & -2+4k & -2 \\
\Lambda_{N=2} & 2 & 0 & 
\end{array}
\eea
The equations of motion obtained by varying the superpotential with respect to 
each field read
\bea
-\frac{\mu_{2k}}{\sqrt{2}} & = & 
\sum_{i=1}^{N_c-1}\frac{\pa a_i}{\pa u_{2k}} m_{i,dy}^2,
\;\;\; k=1, \cdots,N_c-2 \nonu \\
-\frac{\mu_{2(N_c-1)}}{\sqrt{2}} & = & 
\sum_{i=1}^{N_c-1}\frac{\pa a_i}{\pa s_{2(N_c-1)}} m_{i,dy}^2, \nonu \\
-\frac{\lambda}{\sqrt{2}}  & = & 
\sum_{i=1}^{N_c-1}\frac{\pa a_i}{\pa u} m_{i,dy}^2,
\label{mu}
\eea
and 
\bea
a_i m_{i, dy}=0 \;\;\; i=1, \cdots, N_c-1.
\label{am}
\eea
At a generic point in the moduli space, no massless fields appear (
$a_i\neq 0$ for $ i=1, \cdots, N_c-1 $ ) which
implies $m_{i, dy} =0$ by (\ref{am}). Thus $\mu_{2k}, \mu_{2(N_c-1)}$ and
$\lambda$ vanish according to (\ref{mu}).  Then we obtain
the moduli space of vacua of $N=2$ theory. 

On the other hand, we consider a singular point in the moduli space where
$l$ mutually local dyons are massless ( we can choose local coordinates so that the quantum
discriminant factorizes  into linearly independent factors. This
 implies that all branches
intersect transversely ). 
This means that $l$ one cycles
shrink to zero. Then the curve ({\ref{curve}) of genus $2N_c-1$
degenerates  to a curve of  genus $2N_c-2l-1$.
The right hand side of (\ref{curve}) becomes, for $SO(2N_c)$,    
\bea
y_{even}^2=
C^2_{2N_c}(v^2, u_{2k})  - 
\Lambda_{N=2}^{4(N_c-1)} v^4 =\prod_{i=1}^{l}(v^2-p_i^2)^2
\prod_{j=1}^{2N_c -2l}(v^2-q_j^2)
\label{curve1}
\eea
with $p_i$ and $q_j$ distinct. 
A point in the $N=2$ moduli space of vacua is characterized by $p_i$ and $q_j$.
The degeneracy of this curve is checked by explicitly
evaluating both $y_{even}^2$ and $\pa y_{even}^2/ \pa v^2$ 
at the point $v= \pm p_i$, obtaining thus a
zero.
Since $a_i =0$ for $i=1, \cdots, l$  and $a_i\neq 0$ for 
$ i=l+1, \cdots, N_c-1 $,
(\ref{am}) leads to 
\bea 
m_{i, dy}=0, \;\;\;  i = l+1, \cdots , N_c-1
\label{midy}
\eea
while $m_{i, dy} ( i=1, \cdots, l )$ are not constrained.
We will see how the vevs  $m_{i, dy}$ originate from the information about
$N=2$ moduli space of vacua which is encoded in the values of $p_i$ and $q_j$.
We assume that the matrix  $\pa a_i/ \pa u_{2k}$ is nondegenerate and
a complex $2N_c-2l-1$ dimensional moduli space of $N=1$ vacua remains after
perturbation.
In order to calculate $\pa a_i/ \pa u_{2k}$, which appears in the eq. (\ref{mu}),
we need the relation
between $ \pa a_i/ \pa s_{2k}$ and the period integral
on a basis of holomorphic one forms on the curve\footnote{ We thank A. Hanany for
communicating  us  the misprint of the power of $v$ in the original hep-th version. 
The correct expression appeared in the published version of \cite{hanany}. }, 
\bea
\frac{\pa a_i}{\pa s_{2k}}= \oint_{\alpha_i} \frac{v^{2(N_c-k)}dv}{y}.
\label{as}
\eea
By plugging the expression of $y$ of (\ref{curve1})  
into (\ref{as}) and by integrating along one cycles
around $v=\pm p_i ( i=1, \cdots, l )$, we get 
\bea
\frac{\pa a_i}{\pa s_{2k}} = \frac{p_i^{2(N_c-k)}}
{ \prod_{j \neq i}^l (p_i^2-p_j^2) \prod_t^{2N_c-2l}  (p_i^2-q_t^2)^{1/2} },
\label{as1}
\eea
since the $l$ one cycles shrink to zero.
Through the eqs. (\ref{su}), (\ref{mu}) and (\ref{as1}),
we arrive at the following relation between the parameters $\mu_{2k}$ and the dyon
vevs $m_{i,dy}^2$
\bea
\frac{-\mu_{2k}}{\sqrt{2}} = 
\sum_{i=1}^l \sum_{j=1}^{N_c} \frac{-1}{2k} s_{2(j-k)} p_i^{2(N_c-j)}
\frac{m_{i,dy}^2}{ 
\prod_{s \neq i}^l (p_i^2-p_s^2) \prod_t^{2N_c-2l} (p_i^2-q_t^2)^{1/2} }.
\eea
By using the definition
\bea
\omega_i = \frac{\sqrt{2} m_{i,dy}^2}{
\prod_{s \neq i}^l (p_i^2-p_s^2) \prod_t^{2N_c-2l} (p_i^2-q_t^2)^{1/2} },
\eea
which will be useful for comparison with brane configuration,
we can express the generating function for the $\mu_{2k}$, 
$\sum_{k=1}^{N_c-1} 2k \mu_{2k} v^{2(k-1)}$, in terms of $\omega_i$
as follows:
\bea
\sum_{k=1}^{N_c-1} 2k \mu_{2k} v^{2(k-1)} & = & \sum_{k=1}^{N_c-1} \sum_{i=1}^l 
\sum_{j=1}^{N_c} v^{2(k-1)} s_{2(j-k)} p_i^{2(N_c-j)} \omega_i \nonu \\
& = & 
\sum_{k=-\infty}^{N_c-1} 
\sum_{i=1}^l 
\sum_{j=1}^{N_c} v^{2(k-1)} s_{2(j-k)} p_i^{2(N_c-j)} \omega_i  + 
{\cal O}(v^{-4}) \nonu \\
& = &  \sum_{i=1}^l  \frac{ C_{2N_c}(v^2)}{v^2 (v^2-p_i^2)} \omega_i 
+ {\cal O}(v^{-4}).
\label{mu2k}
\eea
In principle, we can find the parameter $\mu_{2k}$ by reading the right hand side
of (\ref{mu2k}).
This result determines whether a point in the $N=2$ moduli space of vacua
classified by the set of $p_i, q_j$ in (\ref{curve1}) remains as an $N=1$ vacuum
after the perturbation, if given a set of perturbation parameters 
$\mu_{2k}, \mu_{2(N_c-1)}$ and $\lambda$, and gives the dyon vevs $m_{i,dy}^2$.
We will see in section 3 that this corresponds to one of the boundary conditions on
a complex coordinate in $( x^8, x^9 )$ directions as 
$v$ ( which is realized as a complex coordinate 
in $( x^4, x^5 )$ directions in string/M theory point of view ) goes to infinity.
In order to make the comparison with the brane picture, 
it is very useful to define the polynomial
$H_{even}(v^2)$ of degree $2l-4$ by:
\bea
\sum_{i=1}^l \frac{\omega_i}{v^2(v^2-p_i^2)} = 
\frac{2H_{even} (v^2)}{ \prod_{i=1}^l (v^2-p_i^2)}.
\label{heven}
\eea
At a given point $p_i$ and  $q_j$ in the $N=2$ moduli space of vacua, 
$H_{even} (v^2)$ determines the dyon vevs, that is,
\bea
m_{i,dy}^2 = \sqrt{2} p_i^2 H_{even} (p_i^2) \prod_m (p_i^2-q_m^2)^{1/2},
\label{dyoneven}
\eea
which will be described in terms of the geometric brane picture in next section.
Therefore, all the vevs of dyons $m_{i, dy} ( i=1, \cdots, l )$ are found:
$m_{i, dy} ( i=l+1, \cdots, N_c-1 )$ vanishes according to (\ref{midy}).

Similarly we can proceed \footnote{
We will describe $SO(2N_c+1)$ case very briefly and simply write down the main results
throughout this paper because the arguments for $SO(2N_c)$ case go through in the
same way. } 
the case of $SO(2N_c+1)$ by starting from
\bea
C^2_{2N_c}(v^2, u_{2k})  - 
\Lambda_{N=2}^{4N_c-2} v^2 =\prod_{i=1}^{l}(v^2-p_i^2)^2
\prod_{j=1}^{2N_c -2l}(v^2-q_j^2)
\eea going near a singular point where we get
\bea
\sum_{k=1}^{N_c} 2k \mu_{2k} v^{2(k-1)}  & = &
 \sum_{k=1}^{N_c} \sum_{i=1}^l 
\sum_{j=1}^{N_c} v^{2(k-1)} s_{2(j-k)} p_i^{2(N_c-j)} \omega_i \nonu \\
& = &  \sum_{i=1}^l   \frac{C_{2N_c}(v^2)}{(v^2-p_i^2)} \omega_i 
+ {\cal O}(v^{-2}).
\label{muodd}
\eea
We will see how this generating function arises when we consider the brane approach which
is the main subject in section 3.
Also we define the polynomial
$H_{odd}(v^2)$ of degree $2l-2$ by:
\bea
\sum_{i=1}^l \frac{\omega_i}{(v^2-p_i^2)} = 
\frac{2H_{odd} (v^2)}{ \prod_{i=1}^l (v^2-p_i^2)},
\label{hodd}
\eea
which will be of use later.
At a given point $p_i$ and $q_m$, $H_{odd} (v^2)$ determines the dyon vevs
\bea
m_{i,dy}^2 = \sqrt{2}  H_{odd} (p_i^2) \prod_m (p_i^2-q_m^2)^{1/2}.
\eea
Of course, the equations of motion leads to
$m_{i, dy} =0$ for $i=l+1, \cdots, N_c-1$.

$\bullet$ Yang-Mills Theory with Massless Matter

The theory with $N_f$ flavors is similar to the pure case we have discussed.
When some of branch points of (\ref{curve}) collide as we change the
moduli, the Riemann surface degenerates and gives a singularity
in the effective theory corresponding to an additional massless field.
The gauge group is $SO(r) \times U(1)^{(N_c-r)/2}$ where $N_c-r$ is even. 
At very special points in the moduli space, 
there are $(N_c-r)/2$ hypermultiplets charged under these $U(1)$'s
becoming simultaneously massless.
The superpotential at these points is  
\bea
W_{even}= \sqrt{2} \sum_{i=1}^{N_c-r-1} M_i A_i M_i + \sum_{k=1}^{N_c-2}
\mu_{2k} U_{2k} +\mu_{2(N_c-1)} S_{2(N_c-1)}+\lambda U
\eea
and 
\bea
W_{odd}= \sqrt{2} \sum_{i=1}^{N_c-r-1} M_i A_i M_i + \sum_{k=1}^{N_c-1}
\mu_{2k} U_{2k}+ \mu_{2N_c} S_{2N_c}.
\eea
As we did in the pure case, the equations of motion can be written as
\bea
-\frac{\mu_{2k}}{\sqrt{2}}
-\sum_{j=N_c-r}^{N_c-2} \frac{\mu_{2j}}{\sqrt{2}} \frac{\pa u_{2j}}
{\pa u_{2k}}
& = & 
\sum_{i=1}^{N_c-r-1}\frac{\pa a_i}{\pa u_{2k}} m_{i,dy}^2,
\;\;\; k=1, \cdots,N_c-r-2 \nonu \\
-\frac{\mu_{2(N_c-1)}}{\sqrt{2}} & = & 
\sum_{i=1}^{N_c-r-1}\frac{\pa a_i}{\pa s_{2(N_c-1)}} m_{i,dy}^2, \nonu \\
-\frac{\lambda}{\sqrt{2}} & = & 
\sum_{i=1}^{N_c-r-1}\frac{\pa a_i}{\pa u} m_{i,dy}^2,
\label{mu1}
\eea
and
\bea
a_i m_{i, dy}=0 \;\;\; i=1, \cdots, N_c-r-1.
\label{am2}
\eea
Notice that the extra term in the left hand side of (\ref{mu1}) comes from
the fact that $U_{2k}$ for $k > N_c-r-1$ are dependent on $U_{2k}$ for
$k \leq N_c-r$.
At a generic point in the moduli space, no massless fields $m_{i, dy}$ appear
($a_i\neq 0$ for $ i=1, \cdots, N_c-r-1 $ ) which
implies $m_{i, dy} =0$ by (\ref{am2}).  Then we get the moduli space of vacua
of $N=2$ theory since all the parameters are zero which gives $\Delta W=0$.

In order to reduce this case to the one analogous to the pure Yang-Mills case where
mutually local dyons are massless,
we define $\tilde{y}^{2}_{even} = y^{2}_{even}/v^{4r}$ to get the $2r$ branch points
of Riemann surface:
the curve ({\ref{curve}) of genus $2N_c-2r-1$
degenerates  to a curve of  genus $2N_c-2r-2l-1$. Then
\bea
\tilde{y}_{even}^2 = C^2_{2(N_c-r)}(v^2)  - \Lambda_{N=2}^{2(2N_c-2-N_f)}
v^{2N_f-4r+4} 
\label{evenmatter}
\eea 
or
\bea
\tilde{y}_{even}^2 & = &
C^2_{2(N_c-r)}(v^2)  -   \Lambda_{N=2}^{4(N_c-r)-(2N_f-4r+4)}
v^{2N_f-4r+4} \nonu \\
& = &
\prod_{i=1}^{l}(v^2-p_i^2)^2
\prod_{j=1}^{2(N_c-r  -l)}(v^2-q_j^2)
\eea
with all $p_{i}, q_{j}$ distinct. Similarly, by redefinition
of  
$\tilde{y}^{2}_{odd} = y^{2}_{odd}/v^{4r+2}$ we obtain
\bea
\tilde{y}_{odd}^2 & = &
C^2_{(2N_c-2r-1)}(v^2) -   \Lambda_{N=2}^{2(2N_c-2r-1)-(2N_f-4r)}
v^{2N_f-4r} \nonu \\
& = &
\prod_{i=1}^{l}(v^2-p_i^2)^2
\prod_{j=1}^{2(N_c-r  -l)}(v^2-q_j^2).
\eea
Now the equation (\ref{am2}) implies that
\bea 
m_{i, dy}=0, \;\;\; i = l+1, \cdots, N_c-r-1
\eea
while $m_{i, dy}$ for $i=1, \cdots , l$ are not constrained since
$a_i=0$ for $i=1, \cdots, l$ and $a_i \neq 0 $ for $i= l+1, \cdots, N_c-r-1$.
By analogy with the pure Yang-Mills case, we use again the relation:
\bea
\frac{\partial a_i}{\partial s_{2k}}=
\oint_{\alpha_i} \frac{v^{2(N_c-k)}dv}{{\tilde{y}}_{even}}
= \oint_{\alpha_i} \frac{v^{2(N_c-r-k)}dv}{y_{even}}.
\eea
Then, after making the steps between (\ref{as})-(\ref{mu2k}),
it turns out the generating function for $\mu_{2k}$ is given by:
\bea
\sum_{k=1}^{N_c-1} 2k \mu_{2k} v^{2(k-1)}
=  \sum_{i=1}^l  \frac{ C_{2(N_c-r)}(v^2)}{v^2(v^2-p_i^2)} \omega_i 
+ {\cal O}(v^{-4})
\label{matteven}
\eea
where the relation between the function $H(v^2), \omega_{i}$ and the dyon vevs 
$m_{i, dy}^2$ is the same as in pure Yang-Mills case.
Also we get similar result for $SO(2N_c+1)$,
\bea
\sum_{k=1}^{N_c} 2k \mu_{2k} v^{2(k-1)}
=  \sum_{i=1}^l  \frac{ C_{2N_c-2r-1}(v^2)}{(v^2-p_i^2)} \omega_i 
+ {\cal O}(v^{-2}).
\label{mattodd}
\eea

\subsection{The Meson Vevs}

Let us discuss the vevs of the meson field along the singular locus of the Coulomb
branch. This is due to the nonperturbative effects of $N=1$ theory and obviously
was zero before the perturbation (\ref{delw}). We will see the property of exactness
in field theory analysis in the context of M theory fivebrane in section 4. Equivalently,
the exactness of superpotential for any values of the parameters is to assume 
$W_{\Delta}=0$.

$\bullet$ $SO(2N_c)$ case

We will follow the method presented in \cite{efgir}.
Let us consider the vacuum where one massless dyon exists 
with unbroken $SU(2) \times U(1)^{N_c-1}$ where
\bea
\Phi_{even}^{cl}= \sigma_2 \otimes 
\mbox{diag} (a_1, a_1, a_2, \cdots, a_{N_c-1}),
\;\;\; \sigma_2=
\left( \begin{array}{cc}
0 & -i \\  
i & 0  
\end{array} \right)
\label{vacuaeven}
\eea
and the chiral multiplet $Q=0$.
These eigenvalues of $\Phi$ can be obtained by differentiating the superpotential
(\ref{superpotential}) with respect to $\Phi$ and setting the chiral multiplet $Q=0$.
\bea
W'(\Phi) & = & \sum_{i=1}^{N_c-2} 2i \mu_{2i} (\Phi^{2i-1})_{ij} +
\mu_{2(N_c-1)} \frac{\pa s_{2(N_c-1)}}{\pa \Phi} \nonu \\
 & & -
\frac{\lambda}{2^{N_c} (N_c-1)!} \epsilon_{ijk_1l_1 \cdots k_{N_c}l_{N_c}}
\Phi^{k_1l_1} \Phi^{k_2l_2} \cdots \Phi^{k_{N_c}l_{N_c}} =0.
\eea
The vacua with classical $SU(2) \times U(1)^{N_c-1}$ group are those with
two eigenvalues equal to $a_1$ and the rest given by $a_2, a_3, \cdots, a_{N_c-1}$.
It is known from \cite{ty,kty} that, if using $s_{2(N_c-1)}$ in the superpotential
perturbation rather than $u_{2(N_c-1)}$
the degenerate eigenvalue of $\Phi$
is obtained to be:
\bea
a_1^2=\frac{(N_c-2) \mu_{2(N_c-2)}}{(N_c-1) \mu_{2(N_c-1)}}.
\label{a1}
\eea
We will see in section 3 that in the context of string/M theory,
the asymptotic behavior of a complex coordinate in $( x^8, x^9 )$ directions for large $v$
determines this degenerate eigenvalue by using the condition for generating function of
$\mu_{2k}$ (\ref{mu2k}).
Recall that  $\mu_{2(N_c-1)} s_{2(N_c-1)}$ term in (\ref{delw}) is used rather than
$\mu_{2(N_c-1)} u_{2(N_c-1)}$ to get this result.
The scale matching condition between the high energy $SO(2N_c)$ scale
$\Lambda_{N=2}$ and the low energy $SU(2)$ scale $\Lambda_{SU(2), N_{f}}$
is related by the following relation
\bea
\Lambda_{SU(2), N_f}^{6-2N_f}=(2(N_c-1) \mu_{2(N_c-1)})^2 \Lambda_{N=2}^
{4(N_c-1)-2N_f}.
\eea
After integrating out $SU(2)$ quarks we obtain the scale matching
between $\Lambda_{N=2}$ and $\Lambda_{SU(2)}$ for pure $N=1$ $SU(2)$
gauge theory. That is,
\bea
\Lambda_{SU(2)}^6=(2(N_c-1) \mu_{2(N_c-1)})^2 \Lambda_{N=2}^
{4(N_c-1)-2N_f} \mbox{det} (a_1^2-m^2)
\eea 
where matrix $a_1^2$ means
$i \sigma \otimes a_1^2$ and quark mass matrix $m$ being 
$( { 0 \atop 1 }{  1 \atop 0 }  ) 
\otimes \mbox{diag} ( m_{1}, \cdots, m_{N_f} ) $.
Then the full exact low energy effective superpotential is given by
\bea
W_L & = & W_{cl} \pm 2 \Lambda_{SU(2)}^3 \nonu \\
&  = &
 \sum_{k=1}^{N_c-2} \mu_{2k} \mbox{Tr} (\Phi_{cl}^{2k}) +
\mu_{2(N_c-1)} s^{cl}_{2(N_c-1)}+
\lambda \mbox{Pf} \Phi_{cl} \pm 2 \Lambda_{SU(2)}^3
\label{lowsuperpo}
\eea
where
$W_{cl}$ is the superpotential evaluated in the classical
$SU(2) \times U(1)^{N_c-1}$ vacua and the last term is generated by
gaugino condensation in the low energy $SU(2)$ theory ( the sign reflects the vacuum
degeneracy ).
In terms of the original $N=2$ scale, it is written as
\bea
W_{even} (\mu_{2k}, \lambda, m) 
& = & \sum_{k=1}^{N_c-2} \mu_{2k} \mbox{Tr} (\Phi_{cl}^{2k}) +
\lambda \mbox{Pf} \Phi_{cl} 
+\mu_{2(N_c-1)} s^{cl}_{2(N_c-1)}
\nonu \\
& & \pm 4(N_c-1) \mu_{2(N_c-1)} \Lambda_{N=2}^
{2(N_c-1)-N_f} \mbox{det} (a_1^2-m^2)^{1/2}.
\eea
Therefore,
one can obtain the vevs of meson
$M_i=Q^i_a Q^i_a$ by taking the mass matrix $m$ to be
$( { 0 \atop 1 }{  1 \atop 0 }  ) 
\otimes \mbox{diag} ( m_{1}, \cdots, m_{N_f} ) $ which gives
\bea
M_i=\frac{\pa W_{even}}{\pa m_i^2}=\pm \frac{4 \Lambda_{N=2}^
{2(N_c-1)-N_f}}{\sqrt{2}(a_1^2-m_i^2)}
(N_c-1) \mu_{2(N_c-1)}  \mbox{det} (a_1^2-m^2)^{1/2},
\label{meson}
\eea
where $a_1$ is given by ({\ref{a1}).
It is easy to see that the vacua of gauge invariant order parameters which are
obtained from $W_L$ parameterize the singularities of the curve (\ref{curve}) and
reproduce the $N=2$ curve (\ref{curve}).
We will see in section 3 that the finite value of a complex coordinate 
in $( x^8, x^9 )$ directions corresponds to the above vevs of meson
when $v \rightarrow \pm m_i$ and the other complex coordinate in $( x^6, x^{10} )$ 
directions vanishes.

$\bullet$ $SO(2N_c+1)$ case

Let us go now to the $SO(2N_{c}+1)$ group for which there is no contribution from
$\mbox{Pf} \Phi$ and consider again the case of one
massless dyon, i.e., the case of unbroken $SU(2)\times U(1)^{N_{c}-1}$ vacua
with:
\bea
\Phi_{odd}^{cl}= \sigma_2 \otimes 
\mbox{diag} (a_1, a_1, a_2, \cdots, a_{N_c-1}, 0),
\eea
which can be determined by differentiating the superpotential with respect to $\Phi$,
\bea
W'(\Phi)=\sum_{i=1}^{N_c-1} 2i \mu_{2i} (\Phi^{2i-1})_{ij}+
\mu_{2N_c} \frac{\pa s_{2N_c}}{\pa \Phi} =0.
\eea
Again we use the result of \cite{ty,kty} where the degenerate eigenvalue of $\Phi$
was obtained to be
\bea
a_1^2=\frac{(N_c-1) \mu_{2(N_c-1)}}{N_c \mu_{2N_c}}.
\label{a1odd}
\eea
We will use this value when we discuss the property of the function of a complex
coordinate in $( x^8, x^9 )$ directions in section 3.
The scale matching between the high energy $SO(2N_{c}+1)$ scale and the low 
energy $SU(2)$ scale is 
\bea
\Lambda_{SU(2), N_f}^{6-2N_f}=(2N_c \mu_{2N_c})(2(N_c-1) \mu_{2(N_c-1)}) 
\Lambda_{N=2}^
{2(2N_c-1-N_f)}
\eea
and after integrating out $SU(2)$ quarks it leads to for pure $N=1$ $SU(2)$ gauge theory
\bea
\Lambda_{SU(2), N_f}^{6}=(2N_c \mu_{2N_c})(2(N_c-1) \mu_{2(N_c-1)}) 
\Lambda_{N=2}^{2(2N_c-1-N_f)} \mbox{det} (a_1^2-m^2)
\eea
as in $SO(2N_c)$ case. The matrices $a_1$ and $m$ are the same as those 
in even $N_c$ case.
As a result the low energy effective superpotential is given by
\bea
W_L=W_{cl}  \pm 2 \Lambda_{SU(2)}^3=
\sum_{k=1}^{N_c-1} \mu_{2k} \mbox{Tr} (\Phi_{cl}^{2k})+\mu_{2N_c} s^{cl}_{2N_c} 
 \pm 2 \Lambda_{SU(2)}^3,
\label{lowsuperpo1}
\eea
where again the sign reflects the vacuum degeneracy.
The quadratic degeneracy of the curve (\ref{curve}) is confirmed by
the vevs of gauge invariants obtained by $W_L$.
In terms of original $N=2$ scale it is written as
\bea
W_{odd} (\mu_{2k}, m) 
&  = & \sum_{k=1}^{N_c-1} \mu_{2k} \mbox{Tr} (\Phi_{cl}^{2k}) +
\mu_{2N_c} s^{cl}_{N_c} 
\nonu \\
& &  \pm 4 \sqrt{N_c(N_c-1) \mu_{2N_c} \mu_{2(N_c-1)}} \Lambda_{N=2}^
{2N_c-1-N_f} \mbox{det} (a_1^2-m^2)^{1/2}.
\eea
Finally,
one gets the vevs of meson
$M_i=Q^i_a Q^i_a$ by using the corresponding mass matrix $m$,
\bea
M_i=\frac{\pa W_{odd}}{\pa m_i^2}=\pm \frac{4 \Lambda_{N=2}^{2N_c-1-N_f}
}{\sqrt{2}(a_1^2-m_i^2)}
\sqrt{N_c(N_c-1) \mu_{2N_c} \mu_{2(N_c-1)}}  \mbox{det} (a_1^2-m^2)^{1/2},
\label{meson1}
\eea
where $a_1$ given by (\ref{a1odd}).
We will learn how this vevs of meson will occur  and relate to the asymptotic  location
in $ ( x^8, x^9 )$ directions of semiinfinite D4 branes in brane geometry.

\subsection{Several Massless Dyons}

We discuss now the case of several massless dyons.
The basic procedure in this direction already appeared in \cite{kss} 
for the $SU(N_c)$ case.
Let us start with the $SO(2N_{c})$ case. The classical moduli space is given then by
\begin{equation}
\Phi^{cl}_{even} = \sigma_{2}\otimes \mbox{diag}(a_{1}^{r_{1}},\cdots,
a_{k}^{r_{k}}),
\label{several1}
\end{equation}
where the eigenvalue $a_1$ occurs $r_1$ times, the eigenvalue $a_2$ does
$r_2$ times and so on.
When $r_1=2$ and $r_i=1 ( i > 1 )$, this will lead to the case of one dyon (\ref{vacuaeven}).
The unbroken group is identical to the one of $SU$ case, i.e.,
$SU(r_{1})\times \cdots \times SU(r_{k})\times U(1)^{k-1}$. In 
(\ref{several1}), we consider that all of the $a_{i}$'s are non zero. If some
of them are zero, the unbroken gauge group will be a product of several $SO$ groups with
several $SU$ groups. 

The procedure to obtain the meson vevs is already clear from the arguments of
one massless dyon. That is, we decompose
$\Phi = \Phi_{cl} + \delta\Phi$ and some of the $\delta\Phi$'s commute with
$\Phi_{cl}$ and are integrated first out. Then one obtains just a product
of $SU(r_{i})$ groups, each one with adjoints. Some of the vector particles
are massive after the symmetry breaking and are  to be integrated out. 
After that we integrate out the adjoint fields in each $SU(r_i)$ so
we go from $N=2$ to $N=1$ theory.  In the $N=1$ theory we integrate out
the quarks. The final formula for the scale of the pure $N=1$
$SU(r_{i})$ theory is: 
\begin{equation}
\Lambda^{3r_{i}}_i = \mbox{det} (a_{i}^{2} - m^{2}) \phi(a_{i})^{r_{i}}
\prod_{j\ne i} (a_{j}^{2} - a_{i}^{2})^{r_{i}-2r_{j}} \Lambda^{4(N_{c}-1)
- 2N_{f}}_{N=2}
\end{equation}
for some polynomial $\phi(a_i)$
and the  low-energy effective superpotential is given by
\begin{equation}
W_{even} = \sum_{k=1}^{N_{c}-2}\mu_{2k} \mbox{Tr} (\Phi_{cl}^{2k}) + 
 \mu_{2(N_c-1)} s^{cl}_{2(N_c-1)}+
\lambda \mbox{Pf}\Phi_{cl}
\pm \sum_{i=1}^k j_{i}\nu_{i}\Lambda_{i}^{3}
\end{equation}
where $\nu_i$ is an $r_i$-th root of unity
and after differentiating with respect to $m_{j}^2$ we obtain:
\begin{equation}
\label{result1}
\sqrt{2} M_{j} = \pm \sum_{i} \frac{\nu_{i}\phi(a_{i})}{a_{i}^{2}-m_{j}^{2}}
\mbox{det}(a_{i}^{2} - m^{2})^{1/r_{i}} \prod_{j\ne i} (a_{j}^2-a_{i}^2)^{1-
2r_{j}/r_{i}} \Lambda_{N=2}^{\frac{4(N_{c}-1)-2N_{f}}{r_{i}}}.
\end{equation}

In the case of $SO(2N_{c}+1)$ gauge group, the classical moduli space is
taken to be:
\begin{equation}
\Phi^{cl}_{odd} = \sigma_{2} \otimes \mbox{diag} (a_{1}^{r_{1}},\cdots
a_{k}^{r_{k}}, 0).
\end{equation}
Then we obtain the meson vevs by using the same procedure as in the 
even case and we integrate out all the massive fields. The only difference as
compared with the even case (\ref{result1}) is the power of $\Lambda_{N=2}$ 
which becomes $\frac{2(2N_{c} - 1)-2N_f}{r_i}$.
We will find for the unbroken gauge group $SU(2)$ that there exists an agreement
between the meson vevs result in this section and the one which will be discussed
in section 4. However, we will find for the unbroken gauge group $SU(r), r >2$ that
there is a disagreement between these two approaches.

\section{ Brane Configuration from M Theory }
\setcounter{equation}{0}

In this section we study the theory with the superpotential perturbation
$\Delta W$ (\ref{delw}) by analyzing M theory fivebranes. 
Let us first describe them in the type IIA brane configuration.

Following \cite{egk}, the brane
configuration in $N=2$ theory consists of
three kind of branes: the two parallel NS5
branes extend in the directions $(x^0, x^1, x^2, x^3, x^4, x^5)$, the D4 branes
are stretched between two NS5 branes and extend over $(x^0, x^1, x^2, x^3)$ and
are finite in the direction of $x^6$, and the D6 branes extend in the directions
$(x^0, x^1, x^2, x^3, x^7, x^8, x^9)$. 
In order to study 
orthogonal gauge groups, we will consider an O4
orientifold which is parallel to the D4 branes in order to keep the
supersymmetry and is not of finite extent in $x^6$ direction. The D4 branes
is the only brane which is not intersected by this O4 orientifold. 
The orientifold
gives a spacetime reflection as $(x^4, x^5, x^7, x^8, x^9) \rightarrow
(-x^4, -x^5, -x^7, -x^8, -x^9)$, in addition to the gauging of worldsheet
parity $\Omega$. 
The fixed points of the spacetime symmetry define this O4 planes.
Each object which does not lie at the fixed points ( i.e., over the orientifold
plane ), must have its mirror image. Thus NS5 branes have a mirror in $(x^4,
x^5)$ directions and D6 branes have a mirror in $(x^7, x^8, x^9)$ directions.

For $SO(2N_c)$ gauge group, each D4 brane at $v=x^4+i x^5$
has its mirror image at $-v$: $N_c$ D4 branes and its mirror $N_c$ ones.
Similarly, for $SO(2N_c+1)$ gauge group, there exist an extra single
D4 brane which lies over the O4 orientifold being frozen at $v=0$ because
it does not contain its mirror image, as well as 
$N_c$ D4 branes and their $N_c$ mirror branes.
Another important ingredient of O4 orientifold is its charge which is
related to the sign of $\Omega^2$. When the
D4 brane carries one unit of this charge, the charge of the O4 orientifold 
is $\mp 1$, for $\Omega^2= \pm 1$ in the D4  brane sector.
We are considering  the 4 dimensional $N=2$ supersymmetric gauge theory
on D4 brane's worldvolume, $(x^0, x^1, x^2, x^3)$ directions. The Higgs branch of
the theory can be described as the D4 branes broken the D6 brane, suspended
them and being allowed to move on the directions $( x^7, x^8, x^9 )$.
The dimension of the Higgs moduli space is found by
counting all possible breakings of D4 branes into D6 branes.

In order to realize the $N=1$ theory with a perturbation (\ref{delw})
we can think of a single NS5 brane and {\it multiple copies } of NS'5 branes
which are orthogonal to a NS5 brane 
with worldvolume, $(x^0, x^1, x^2, x^3, x^8, x^9)$ and between them there
exist D4 branes intersecting D6 branes. The number of NS'5 branes is
$N_c-2$ for $SO(2N_c)$ and $N_c-1$ for $SO(2N_c+1)$ by identifying the power
of adjoint field appearing in the superpotential (\ref{delw}).
The brane description for $N=1$ theory with a superpotential (\ref{superpotential})
where $\mu_{2(N_c-1)}=\mu_{2N_c}=\lambda=0$
has been studied in the paper \cite{egkrs} in type IIA brane configuration. In this case,
all the couplings, $\mu_{2k}$ can be regarded as tending uniformly to infinity. On the other
hand, we will see in M theory configuration there will be no such restrictions.

\subsection{ M Theory Fivebrane Configuration}

$\bullet$ $SO(2N_c)$ case

Let us describe how the above brane configuration is embedded in 
M theory in terms of
a single M theory fivebrane whose worldvolume is
${\bf R^{1,3}} \times \Sigma$ where $\Sigma$ is identified with Seiberg-Witten
curves \cite{as,hanany,dkp} that determine  the solutions to Coulomb branch 
of the field theory.
As usual, we write $s=(x^6+i x^{10})/R, t=e^{-s}$
where $x^{10}$ is the eleventh coordinate of M theory which is compactified
on a circle of radius $R$. Then the curve $\Sigma$, describing
$N=2$  $SO(N_c)$ gauge theory with $N_f$ flavors and even $N_c$,
is given \cite{lll} by an equation in $(v, t)$ space
\bea
t^2 - \frac{2C_{2N_c}(v^2, u_{2k})}{v^2} t  + \Lambda_{N=2}^{4N_c-4-2N_f}
\prod_{i=1}^{N_f} (v^2 -m_i^2) = 0.
\label{cur}
\eea 
Here  $C_{2N_c}(v^2, u_{2k})$ is a degree $2N_c$ polynomial in $v$ 
with only even degree of terms
and  the coefficients
depending on the moduli $u_{2k}$, and $m_i$ is the mass of quark.
It is easy to check that this description is the same as (\ref{curve})
under the identification
\bea
v^2 t= y +C_{2N_c}(v^2, u_{2k}).
\eea

By adding (\ref{delw}) which corresponds to the adjoint chiral multiplet, 
the $N=2$ supersymmetry  will be broken to $N=1$. 
To describe the corresponding brane configuration in M theory,
let us introduce a complex coordinate
\bea
w=x^8+i x^9.
\eea
To match the superpotential perturbation $\Delta W_{even}$ (\ref{delw}),
we propose the following boundary conditions 
for $SO(2N_c)$
\bea
\begin{array}{ccl}
w^{2} & \rightarrow & \sum_{k=2}^{N_c-1} 2k \mu_{2k} v^{2(k-1)}\;\;\;\mbox{as}~
v \rightarrow \infty,~
t \sim \Lambda_{N=2}^{2(2N_c-2-N_f)}v^{2N_f-2N_c+2},\\
w & \rightarrow & 0\;\;\;\mbox{as}~
v \rightarrow \infty,~
t \sim v^{2N_c-2}. \\
\end{array}
\label{boundary}
\eea
After deformation, $SU(2)_{7,8,9}$ is broken to $U(1)_{8,9}$ if $\mu_{2k}$
has the charges $(4-4k, 4)$ under $U(1)_{4,5} \times U(1)_{8,9}$.
The charges of coordinates and parameters are given by 
\bea
\begin{array}{cccl}
&U(1)_{4,5}&U(1)_{8,9}&\\
v & 2 & 0 & \\
w & 0 & 2 & \\
t & 4(N_c-1) & 0 & \\
\mu_{2k} &4-4k & 4 \\
\Lambda_{N=2} & 2 & 0 &
\end{array}
\eea
where $U(1)_{4,5}=U(1)_R$ and $U(1)_{8,9}=U(1)_J$ we have mentioned last section.
If we consider now only the value $k=2$, this reduces to the case of
\cite{aotsept} and one obtains in (\ref{boundary}) that 
$w^2 \sim \mu_{4} v^{2}$ as $v \rightarrow \infty$ which is
the same as the relation $w \rightarrow \mu v$ obtained in 
\cite{aotsept} if we identify $\mu_{4}$ with $\mu^{2}$. This 
identification comes also from the $U(1)_{4,5}$ and $U(1)_{8,9}$ 
charges of $\mu$ and $\mu_{4}$. So the reduction to the case of a single
NS5 brane is found.
 
After perturbation, only the singular part of the $N=2$ Coulomb branch with
$l$ or more mutually local massless dyons remains in the moduli space of vacua.
The corresponding brane configuration is possible only when
the curve $\Sigma$ degenerates to a curve of  genus
less than $2N_c-2l-1$.
Let us construct the M theory fivebrane configuration corresponding to
the correct boundary conditions and assume
the condition that $w^{2}$ is a rational function of $v^{2}$ and $t$.
Our result is really similar to the case of $SU(N_c)$ \cite{dbo} and we
will follow their notations.
We  write $w^{2}$ as follows
\bea
w^{2}(t,v^2) = \frac{a(v^2) t + b(v^2)}{c(v^2) t + d(v^2)},
\eea
where $a, b, c, d$ are arbitrary polynomials of $v^2$ and $t$ satisfies the eq. 
(\ref{cur}). 
Now we can calculate the following two quantities using the two
solutions of $t$, denoted by $t_+$ and $t_-$ which satisfy
the equation for $t$, (\ref{cur})
\bea
w^{2}(t_+(v^2),v^2)+w^{2}(t_-(v^2),v^2) = \frac{2acG+2adC+2bcC+2bd}{c^2 G + 2 cdC + d^2}
\eea
and
\bea
w^{2}(t_+(v^2),v^2)-w^{2}(t_-(v^2),v^2) = \frac{2(ad-bc)S\sqrt{T}}{v^2(c^2 G +2 cdC + d^2)}
\eea
where  
there is a relation between 
\bea
C \equiv C_{2N_c} (v^2, u_{2k})/v^2 \;\;\; \mbox{and} \;\;\; G \equiv
\Lambda_{N=2}^{4N_c-4-2N_f}
\prod_{i=1}^{N_f} (v^2 -m_i^2)
\eea
implying that
\bea
C^2 -  G(v^2) \equiv \frac{S^2(v^2) T(v^2)}{v^4} 
\eea
where
\bea
S(v^2)=\prod_{i=1}^{l} (v^2-p_i^2), \;\;\;
T(v^2)=\prod_{j=1}^{2N_c-2l} (v^2-q_j^2)
\eea
with all $p_i, q_j$'s different. Remember that $N=2$ moduli space of vacua
is determined by these $p_i$ and $q_j$.
Since $w^{2}$ has no poles for finite value of $v^2$,
$w^{2}(t_+(v^2),v^2) \pm w^{2}(t_-(v^2),v^2)$ also does not have poles which leads to
arbitrary polynomials $H(v^2)$ and $N(v^2)$ given by
\bea 
\frac{acG +  adC + bcC +bd}{c^2 G + 2  cdC + d^2} & = &  N,  \\
\frac{(ad-bc)S}{v^2(c^2 G +2 cdC + d^2)} & = & H.
\eea
It will turn out that the function $H(v^2)$ is exactly 
the same as the one (\ref{heven}) or (\ref{hodd})
defined in field theory analysis.
By making a shift of $a \rightarrow a +Nc, b \rightarrow b +Nd$ due to the 
arbitrariness of the polynomials $a$ and $b$  the following
relations come out.
\bea 
\nonumber
w^{2} & = & N +   \frac{a(v^2) t + b(v^2)}{c(v^2) t + d(v^2)},  \\
\nonu
0 & = & acG +  adC+  bcC+bd,  \\
H & = & \frac{(ad-bc)S}{v^2(c^2 G + 2 cdC + d^2)}. 
\label{woh}
\eea
The second equation implies
\bea
a(cG+  dC) + b(d+  cC) = 0 
\eea
which  can be written as
\bea 
cG+dC=-be, \qquad d+cC=ae 
\eea
for arbitrary rational function $e$.
Plugging the values of $c$ and $d$ into the (\ref{woh}), it turns out
$e=S/(H v^2)$. By combining all the information for $b$ and $d$, we get
the most general rational function $w^2$ which has no poles for finite
value of $v^2$ is 
\bea 
w^{2} = N + \frac{at + cHST/v^2 - aC}{ct - c C + aS/(H v^2)}
\eea
where $N, a, c, H$ are arbitrary polynomials.
As we choose two $w^2$'s , each of them possessing different polynomials $a$ and $c$
and subtract them, the numerator of it will be proportional to $t^2-2 C t+G$ which
vanishes according to (\ref{cur}). This means $w^2$ does not depend on $a$ and $c$.
Therefore,
when $c=0$, the form of $w^2$
is very simple. That is,
\bea 
w^{2}  =  N + \frac{v^2 H}{S}(t-C_{2N_c}/v^2).
\label{solw}
\eea
This result will be used throughout  the remaining part of this paper.
Now we want to impose the boundary conditions on $w^2$ from the most general solution 
(\ref{solw}).  From the previous relation, by recognizing $T^{1/2}=v^2 (t-C_{2N_c}/v^2)/S$,
\bea 
w^{2}(t_{\pm}(v^2),v^2) = N \pm H\sqrt{T}
\label{nht}
\eea
and from the boundary condition 
$ w \rightarrow 0 $ for  $v \rightarrow \infty,~
t = t_-(v) \sim v^{2N_c-2}$
it is easy to see the value of $N(v^2)$, 
\bea 
N(v^2) = \left[  H(v^2) \sqrt{T(v^2)} \right]_+ 
\label{n}
\eea
where $\left[ H(v^2) \sqrt{T(v^2)} \right]_+$ means only nonnegative power of $v^2$
when we expand around $v = \infty$.
Next, by applying the other boundary condition 
$ v \rightarrow \infty,~
t \sim \Lambda_{N=2}^{2(2N_c-2-N_f)}v^{2N_f-2N_c+2}$, 
we obtain 
\bea
w^{2} = \left[2  H(v^2) \sqrt{T(v^2)} \right]_+ + {\cal O}(v^{-2}).
\eea
By noting that $w^{2}$ satisfies the following eq.
\bea 
w^4 -2 N w^2 + N^2 - T H^2 = 0,
\label{solofw}
\eea
and restricting the form of $N, T$ and $H$ like as $N \sim c_{1}v^2 + c_{2}, 
T \sim c_{3} v^6 + c_{4} v^{4} + c_{5} v^{2} + c_{6},
H \sim \frac{c_{7}}{v^{2}}$, it leads to
\bea
w^4 +  (c_8 + c_9 v^2) w^2 +c_{10} =0
\eea
for some constants $c_{i} ( i = 1,\cdots, 10 )$.
Then we can solve for $v^2$ in terms of $w^2$ to reproduce the result of
\cite{aotsept}.
As all the couplings $\mu_{2k}$ are becoming very large, $H(v^2)$ and $N(v^2)$
go to infinity. From (\ref{solofw}) $N^2-T H^2$ goes to zero as we take the limit
of $\Lambda_{N=2} \rightarrow 0$. This tells us that $w^2$ becomes
$\frac{N^2-T H^2}{2N}$ and as $N(v^2)$ goes to zero, $w^2 \rightarrow \infty$
showing the findings in \cite{egkrs}.

The brane configuration was constructed only at the singular
point in the $N=2$ moduli space of vacua where $(v, t)-$plane  curves are
degenerate to curves of  genus $2N_c-2l-1$  given in (\ref{curve1}).
The general solution for $w^2$ is 
\bea 
w^{2} = N(v^2) + v^2 H(v^2) \frac{t-C_{2N_c}(v^2)/v^2}{\prod_{i=1}^{l}
(v^2-p_i^2)} 
\eea
where $H(v^2)$ and $N(v^2)$ are arbitrary polynomials of $v^2$.
The boundary condition determines $N(v^{2})$ as follows
\bea
N(v^2)=\left[  H(v^2) \prod_{j=1}^{2N_c-2l}(v^2-q_j^2)^{1/2} \right]_+.
\eea 
The other boundary condition shows that $w^2$ behaves
as
$ w^{2} \rightarrow \sum_{k=1}^{N_c-1} 2k \mu_{2k} v^{2(k-1)} $ from (\ref{boundary}).
Then by expanding $w^2$ in powers of $v^2$ we can identify $H(v^2)$ with
parameter $\mu_{2k}$. Using 
$T^{1/2}=v^2 (t-C/v^2)/S$ and $t=2C/v^2+ \cdots$ from (\ref{cur}) we get
\bea
w^{2} = 2 v^2 H(v^2) \frac{C_{2N_c}(v^2)/v^2}{\prod_{i=1}^{l}(v^2-p_i^2) }
+ {\cal O}(v^{-2}) =  
\sum_{i=1}^l \frac{ C_{2N_c} (v^2) \omega_i}{v^2(v^2-p_i^2)} =
\sum_{k=1}^{N_c-1} 2k \mu_{2k} v^{2(k-1)}
\eea
where we used the definition of $H_{even}$ in (\ref{heven})
and the generating function of $\mu_{2k}$ in (\ref{mu2k}). From this result
one can find the explicit form of $H(v^2)$ in terms of $\mu_{2k}$ by comparing both sides
in the above relation. This is an explanation for field theory results 
of (\ref{mu2k}) and (\ref{heven}) which determine the $N=1$ moduli space of vacua
after the perturbation, from the point of view of M theory fivebrane. 
It reproduces the equations
which determine the vevs of massless dyons along the singular locus.
The dyon vevs  $m_{i, dy}^2$, given by (\ref{dyoneven})
\bea
m_{i,dy}^2 = \sqrt{2} p_i^2 H_{even} (p_i^2) \sqrt{T(p_i^2)},
\eea
are nothing but the difference between the two finite values of $v^2 w^2$.
This can be seen by taking $v=\pm p_i$ in 
(\ref{nht}) and (\ref{n}).
The $N=2$ curve of (\ref{cur}) and (\ref{curve1}) contains  double points
 at $v=\pm p_i$
and $t=C_{2N_c}(p_i^2)$. The perturbation $\Delta W$ of  (\ref{delw}) splits
 these into
 separate points in $(v, t, w)$ space and the difference in $v^2w^2$ between
 these points  becomes the dyon vevs.
 This is a geometric interpretation of dyon vevs
in M theory brane configuration.

$\bullet$ $SO(2N_c+1)$ case

The curve $\Sigma$ for $N=2$ $SO(2N_c+1)$ gauge theory with $N_f$ flavors
reads in $(v, t)$ space
\bea
t^2 - \frac{2C_{2N_c}(v^2, u_{2k})}{v} t  + \Lambda_{N=2}^{4N_c-2-2N_f}
\prod_{i=1}^{N_f} (v^2 -m_i^2) = 0,
\eea 
where $t$ is related to $y$ by
\bea
v t= y +C_{2N_c}(v^2, u_{2k}).
\eea
The configuration of M theory fivebrane corresponding to type IIA brane configuration
has the following boundary conditions
\bea
\begin{array}{ccl}
w^{2} & \rightarrow & \sum_{k=1}^{N_c} 2k \mu_{2k} v^{2(k-1)}\;\;\;\mbox{as}~
v \rightarrow \infty,~
t \sim \Lambda_{N=2}^{2(2N_c-1-N_f)}v^{2N_f-2N_c+1},\\
w & \rightarrow & 0\;\;\;\mbox{as}~
v \rightarrow \infty,~
t \sim v^{2N_c-1}.\\
\end{array}
\label{boundary1}
\eea
After doing the similar procedure as in $SO(2N_c)$ case, we arrive at the final
expression
\bea
w^{2} = 2 v H(v^2)  \frac{C_{2N_c}(v^2)/v}{\prod_{i=1}^{l}(v^2-p_i^2) }
+ {\cal O}(v^{-2}) =  
\sum_{i=1}^l \frac{ C_{2N_c} (v^2) \omega_i}{(v^2-p_i^2)} = 
\sum_{k=1}^{N_c} 2k \mu_{2k} v^{2(k-1)}
\eea
where we used the definition of $H_{odd}$ in (\ref{hodd})
and the generating function of $\mu_{2k}$ in (\ref{muodd}). 
One can find the explicit form of $H(v^2)$ in terms of $\mu_{2k}$ by comparing both sides. 

\subsection{ Yang-Mills Theory with Massless Matter}

We have seen that 
$( v, \tilde{t}= \tilde{y}/v^2+C_{2(N_c-r)} (v^2, u_{2k})/v^2 )$ curve ({\ref{curve}) of genus 
$2N_c-2r-1$ degenerates to
a curve of  genus $2N_c-2r-2l-1$
by redefining $\tilde{y}^{2}_{even} = y^{2}_{even}/v^{4r}$ to get the $2r$ branches of the curve.
Now it is straightforward to get the most general form of the solution by looking at the
eq. of (\ref{solw}),
\bea 
w^{2} = N(v^2) + v^2 H(v^2) \frac{\tilde{t}-C_{2(N_c-r)}(v^2)/v^2}{\prod_{i=1}^{l}
(v^2-p_i^2)} 
\eea
where $H(v^2)$ and $N(v^2)$ are arbitrary polynomials of $v^2$.
Once again the boundary condition $w \rightarrow 0$ as $v \rightarrow \infty$
and $\tilde{t} \sim \Lambda_{N=2}^{2(2N_c-2-r)} v^{-2N_c+2r+2}$
gives the form of $N(v^2)$
\bea
N(v^2)=\left[  H(v^2) \prod_{j=1}^{2(N_c-r-l)}(v^2-q_j^2)^{1/2} \right]_+.
\eea 
The other boundary condition $ w^{2} \rightarrow
\sum_{k=1}^{N_c-1} 2k \mu_{2k} v^{2(k-1)} $ as $v \rightarrow \infty, \tilde{t} \sim
v^{2(N_c-r-1)}$ and the relation $\tilde{t}= 2C_{2(N_c-r)} + \cdots$ yield to
\bea
w_{even}^{2} & = & 2 v^2 H(v^2) \frac{C_{2(N_c-r)}(v^2)/v^2}{\prod_{i=1}^{l}(v^2-p_i^2) }
+ {\cal O}(v^{-2}) =  
\sum_{i=1}^l \frac{ C_{2(N_c-r)} (v^2) }{v^2(v^2-p_i^2)} \omega_i \nonu \\
& = &
\sum_{k=1}^{N_c-1} 2k \mu_{2k} v^{2(k-1)}
\eea
which is precisely in agreement with the eq. of (\ref{matteven}) which determines
the relation between the function $H(v^2), \omega_i$ and dyon vevs $m_{i,dy}^2$
after perturbation.
Also we will see the $SO(2N_c+1)$ result analogous to the $SO(2N_c)$.
For most general deformation of the brane, it is
\bea 
w^{2} = N(v^2) + v H(v^2) \frac{\tilde{t}-C_{(2N_c-2r-1)}(v^2)/v}{\prod_{i=1}^{l}
(v^2-p_i^2)} 
\eea
and
\bea
N(v^2)=\left[  H(v^2) \prod_{j=1}^{2(N_c-r-l)}(v^2-q_j^2)^{1/2} \right]_+.
\eea 
By applying the second boundary condition, we arrive at
\bea
w_{odd}^{2} & = & 2 v H(v^2) \frac{C_{(2N_c-2r-1)}(v^2)/v}{\prod_{i=1}^{l}(v^2-p_i^2) }
+ {\cal O}(v^{-2}) =  
\sum_{i=1}^l \frac{ C_{2N_c-2r-1} (v^2) }{(v^2-p_i^2)} \omega_i \nonu \\
&  = &
\sum_{k=1}^{N_c} 2k \mu_{2k} v^{2(k-1)}
\eea
which is again the same as the eq. of (\ref{mattodd}).

\section{Brane Configuration and Field Theory  }
\setcounter{equation}{0}

In this section we continue to study for the meson vevs from the singularity structure
of $N=2$ Riemann surface. The vevs of meson will depend on the moduli structure
of $N=2$ Coulomb branch ( See, for example, (\ref{wi2}) ).
Also, the finite values of $w^2$ can be determined fully by using the property of
boundary conditions of $w^2$ when $v$ goes to be very large.
We will illustrate some examples which studied before in field theory analysis that 
was limited for the case of
a small number of $N_c$ because it becomes very difficult to find the vacua
from the quantum discriminant when $N_c$ is very large.

\subsection{ $SO(2N_c)$ Case }

Let us consider the case of finite $w^2$ at $t=0, v=\pm m_i$ and we want 
to compare with the meson
vevs we have studied in (\ref{meson}).
At a point where there exists a single massless dyon ( in other words,
by putting $l=1$ into (\ref{curve1}) )
and recalling the definition of $T(v^2)$, we have for Yang-Mills with  matter
\bea 
C_{2N_c}^2 (v^2) - \Lambda_{N=2}^{4 N_c-4- 2N_f} v^4 
\prod_{i=1}^{N_f} (v^2-m_i^2) = 
(v^2-p_1^2)^2 T(v^2)
\label{onedyon}
\eea
and the function $w^2$ according to (\ref{nht}) and (\ref{n})  reads
\bea
w^2 = \left[ \frac{h}{v^2} \sqrt{T(v^2)} \right]_+ \pm  \frac{h}{v^2} \sqrt{T(v^2)}
\eea
where in this case $l=1$ means that the polynomial $v^2 H(v^2)$ has the degree of zero and
we denote it by a constant $h$. From (\ref{onedyon}) we see
for $N_f < 2N_c-2$
\bea
\frac{\sqrt{T(v^2)}}{v^2} = \frac{C_{2N_c}(v^2)}{v^2(v^2-p_1^2)} + {\cal O}(v^{-4})
\label{Tvv}
\eea
and we decompose $C_{2N_c}$ as
\bea 
\frac{C_{2N_c}(v^2)}{v^2} = \frac{C_{2N_c}(p_1^2)}{p_1^2} + (v^2-p_1^2)\tilde{C}_{2N_c-4}(v^2)
\label{decom}
\eea
for some polynomial  $v^2 \tilde{C}_{2N_c-4}(v^2)$ of degree $2N_c-2$ which
can be determined completely.
This means that the coefficients of $\tilde{C}_{2N_c-4}(v^2)$ can be fixed from 
the explicit form of the polynomial
$C_{2N_c}(v^2)$.
Through (\ref{Tvv}) and (\ref{decom}) the part with nonnegative powers 
of $v^2$ in
$\frac{\sqrt{T(v^2)}}{v^2}$ becomes $\tilde{C}_{2N_c-4}(v^2)$
as follows
\bea
\frac{\sqrt{T(v^2)}}{v^2} = \tilde{C}_{2N_c-4}(v^2) + 
{\cal O}(v^{-2}) \quad \longrightarrow
\quad  \left[ \frac{\sqrt{T(v^2)}}{v^2} \right]_+ = \tilde{C}_{2N_c-4}(v^2).
\eea
Thus as $v \rightarrow \pm m_i$ the finite value of $w^2$, denoted by $w_i^2$ can be written
\bea 
w^2_i = w^2(v^2 \rightarrow m_i^2) = 
h \tilde{C}_{2N_c-4} (m_i^2)  \pm  \frac{h}{m_i^2} \sqrt{T(m_i^2)}.
\label{wi}
\eea From (\ref{onedyon}), the relation
$\frac{\sqrt{T(m_i^2)}}{m_i^2} =  \frac{C_{2N_c}(m_i^2)}{m_i^2(m_i^2-p_1^2)}+ 
{\cal O}(m_i^{-4})$
holds 
and the decomposition of (\ref{decom}) yields the following relation
\bea
\frac{\sqrt{T(m_i^2)}}{m_i^2} =  \frac{ C_{2N_c}(p_1^2)}{p_1^2(m_i^2-p_1^2)} + 
\tilde{C}_{2N_c-4}(m_i^2).
\eea
By plugging this value into (\ref{wi}) and taking the minus sign 
which corresponds to
$t \rightarrow 0$,
we end up with
\bea 
w_i^2 =  \frac{h}{p_1^2} \frac{C_{2N_c}(p_1^2)}{(p_1^2-m_i^2)}.
\label{wi2}
\eea
In order to find $ C_{2N_c}(p_1^2)$ we evaluate it from (\ref{onedyon})
at $v^2=p_1^2$ to arrive at
\bea
w_i^2 = h \Lambda_{N=2}^{ 2N_c-2 - N_f} \frac{\det 
( p_1^2 - m^2)^{1/2}}{(p_1^2-m_i^2)}.
\eea
In the above expression we need to know the values of $h$ and $p_1$.
On the other hand the boundary condition for $w^2$ for large $v$ leads to
\bea
w^2 \sim 2 \frac{h}{v^2} \frac{ C_{2N_c}(v^2)}{v^2-p_1^2} 
\sim 2 h v^{2(N_c-2)} + 2 h p_1^2 v^{2(N_c-3)} + 
\cdots
\eea
which should be equal to
$ \sum_{k=1}^{N_c-1} 2k \mu_{2k} v^{2(k-1)}$.
Then we can read off the values of $h$ and $p_1$ by comparing both sides term by term.
\bea
2h = 2(N_c-1) \mu_{2(N_c-1)} , 
\qquad p_1^2 =\frac{(N_c-2) \mu_{2(N_c-2)}}{(N_c-1) \mu_{2(N_c-1)}}.
\eea
Finally, the finite value for $w^2$ can be written as
\bea
w_i^2 =( N_c-1) \mu_{2(N_c-1)} 
\Lambda_{N=2}^{2N_c-2-N_f}  
\frac{\det ( a_1^2 - m^2)^{1/2}}{(a_1^2-m_i^2)}
\eea
which is exactly, up to constant, the same expression for 
meson vevs (\ref{meson}) obtained from field theory analysis
in the low energy superpotential (\ref{lowsuperpo}). This illustrates the fact that
at vacua with enhanced gauge group $SU(2)$ the effective superpotential by integrating
in method with the assumption of $W_{\Delta}=0$ is really exact.

\underline{ Example 1:  $SO(6)$ with one flavor}

We would like to demonstrate the above descriptions by taking the specific models.
The $N=2$ theory in this model is described by the curve $\Sigma$:
\begin{equation}
t^{2} - \frac{2C_{6}(v^{2}, u_{2k})}{v^{2}} t + \Lambda_{N=2}^{6} (v^{2} - m_1^{2}) = 0
\end{equation}
where the polynomial $C_6 (v^2)$ is given as (\ref{poly})
\begin{equation}
C_{6}(v^{2}, u_{2k}) = v^{6} + s_{2} v^{4} + s_{4} v^{2} + s_{6} 
\end{equation}
in terms of $s$ or 
\begin{equation}
C_{6}(v^{2}, u_{2k}) = v^{6} - \frac{u_{2}}{2}v^{4} - \left(\frac{u_{4}}{4} -
\frac{u_{2}^{2}}{8}\right)v^{2} - \frac{1}{6}\left(u_{6} - \frac{3 u_{4}u_{2}}{4} +
\frac{u_{2}^{3}}{8}\right)
\end{equation}
in terms of $u$.
When one dyon becomes massless the locus in the moduli space becomes:
\begin{equation}
\label{loc1}
C_{6}^{2}(v^{2}, u_{2k}) - \Lambda_{N=2}^{6} v^{4}(v^{2} - m_1^2) = (v^{2} - p_1^{2})^{2}
T(v^2).
\label{exam1}
\end{equation}
By putting  $v^2=p_1^2$ in the
equation (\ref{loc1}) we get one relation and by differentiating  
with respect to $v^{2}$
and evaluating the derived equation at $v^2=p_1^2$ we obtain a second relation
between  
$s_{2}, s_{4}$ and $s_{6}$. So one of those vevs
 will remain undetermined. This is 
because for only one massless dyon there exist  only terms with $v^{4}$ so after 
taking two derivatives, in the right hand side  of (\ref{loc1})
we would have only terms that cannot cancel at
$v^2=p_1^2$. 
For several massless dyons the power of $v^2$ in the right hand side of (\ref{loc1})
would be bigger than $v^{4}$ and we could take two derivatives with respect to
$v^{2}$. Now
we express $s_2$ and $s_4$ in terms of $s_{6}$ whose classical vev vanishes 
\begin{eqnarray}
s_{2} &=& -2p_1^{2} + \frac{s_{6}}{p_1^{4}} \pm 
\frac{\Lambda_{N=2}^{3}}{2b}, \\ \nonumber
s_{4} &=& p_1^{4} - \frac{2s_{6}}{p_1^{2}} \pm 
\frac{\Lambda_{N=2}^{3}}{2b} \sqrt{2m_1^2-p_1^2}
\end{eqnarray} 
where $b=\sqrt{p_1^2-m_1^2}$.
These vevs of gauge invariant variables are exactly the same as
those obtained in \cite{kty} from the low energy effective superpotential, for $N_c=2$,
$ \mu_{2} \mbox{Tr} (\Phi_{cl}^{2})+\mu_{4} s^{cl}_{4} +
\lambda \mbox{Pf} \Phi_{cl}  \pm 2 \Lambda_{SU(2)}^3 $ where
the classical vacua of $ \Phi_{cl}=  \sigma_2 \otimes 
\mbox{diag} (a_1, a_1, a_2) $ breaks $SO(6)$ into $SU(2) \times U(1) \times U(1)$.
For the case of pure Yang-Mills theory, it is known from \cite{bl} that 
only six
points remain as mutually local dyons for $SO(8)$ gauge theory while in $SO(6)$
four points give the correct $N=1$ vacua.
By using the relation (\ref{decom}), it is easy to find the value 
of $N(v^2)$ and from
(\ref{exam1}) one obtains $T(v^2)$.
\bea
N & = & h \left(v^2+\left(p_1^2+s_2\right)-\frac{s_6}{p_1^2 v^2}\right), \nonu \\
N^2 - h^2 T & = & h^2 \left[
-2p_1^4 - 2p_1^2s_2 - 2s_4 - \frac{2s_6}{p_1^2}  \right. \nonu \\
& & + 
 \frac{1}{v^2} \left( -4p_1^6 + \Lambda_{N=2}^6 - 6p_1^4 s_2 - 2p_1^2 s_2^2 - 4p_1^2 s_4 - 
     2s_2 s_4 - 4s_6 - \frac{2s_2 s_6}{p_1^2} \right) \nonu \\
& & +
  \frac{1}{v^4} \left( -5p_1^8 + \Lambda_{N=2}^6 (2p_1^2 -  m_1^2) - 8p_1^6 s_2 - 
     3p_1^4 s_2^2 - 6 p_1^4 s_4 - 4p_1^2 s_2 s_4  \right. \nonu \\
& &  \left. \left. -s_4^2 - 4 p_1^2 s_6 - 2 s_2 s_6 + \frac{s_6^2}{p_1^4}  \right) \right].  
\label{nvalue}
\eea
Also we get the dyon vevs from the relation $m_{1,dy}^4=2 h^2 T(p_1^2)$,
\bea
m_{1, dy}^4 & = & 2 h^2 \left(
15p_1^8 - \Lambda_{N=2}^6 (3p_1^2- m_1^2) + 20 p_1^6 s_2 + 
  6p_1^4 s_2^2 + 12 p_1^4 s_4 + 6 p_1^2 s_2 s_4  \right. \nonu \\
& & \left. + s_4^2 + 6 p_1^2 s_6 + 2 s_2 s_6
 \right).
\eea
The boundary condition for $w^2$ gives the relation between $\mu_4$ and $h$,
\bea
h=2 \mu_4.
\eea
By explicitly calculating $C_6(p_1^2)$, we get the meson vevs,
\bea  
w^2_1= h \frac{\Lambda_{N=2}^3  b}{p_1^2-m_1^2}.
\eea

\underline{ Example 2: $SO(6)$ with two flavors}

Let us consider the previous example with two flavors which 
is described by the curve:
\begin{equation}
t^{2} - \frac{2 C_{6}(v^{2}, u_{2k})}{v^{2}} t + \Lambda_{N=2}^{4} (v^{2} - m_{1}^{2})	
(v^{2} - m_{2}^{2})=0
\end{equation}
where $m_{1}, m_{2}$ are the masses for the two flavors and $C_{6}$ is the
same as before. 
When one dyon becomes massless the locus in the moduli space becomes as in 
(\ref{exam1}):
\begin{equation}
C^{2}_{6} - \Lambda^{4}_{N=2} v^{4} (v^{2} - m_{1}^{2})(v^{2} - m_{2}^{2})
= (v^{2} - p_1^{2})^{2} T(v^2).
\end{equation}  From this equation we again obtain the values for $s_{2k}$. 
As in Example 1, for one
massless dyon we have only two equations for three variables $s_{2}, s_{4}$ and 
$s_{6}$ and we can solve $s_{2}, s_{4}$ in terms  of $s_{6}$. 
Our results are
\bea
s_{2} &=& - 2p_1^{2} + \frac{s_{6}}{p_1^{4}} \pm \frac{\Lambda_{N=2}^{2}}{2B}
 \left( 2p_1^2-m_1^2-m_2^2 \right), \nonu \\ 
s_{4} & = &  p_1^{4}  - \frac{2 s_{6}}{p_1^{2}} \pm \frac{\Lambda_{N=2}^{2}}{2B} 
\left( p_1^2 m_1^2+p_1^2 m_2^2-2 m_1^2 m_2^2 \right)
\label{s24}
\eea
where $B = \pm\sqrt{(p_1^{2} - m_{1}^{2})(p_1^{2} - m_{2}^{2})}$ and we take only 
the plus sign. Remark that $B$ has dimension 2 while $b$  has
dimension 1. 
The value for $N(v^2)$ remains the same as in equation (\ref{nvalue})
given different $s_2$ and $s_4$ as in (\ref{s24}) and from the explicit form of
$T(v^2)$ we obtain the following relation,
\bea
N^2-h^2 T & = &  h^2 \left[ -2p_1^4 + \Lambda_{N=2}^4 - 2p_1^2 s_2 - 2 s_4 - 
\frac{2 s_6}{p_1^2} 
\right. \nonu \\
& & + \frac{1}{v^2}
  \left( -4 p_1^6 + \Lambda_{N=2}^4 ( 2p_1^2 -m_1^2-m_2^2) - 
     6 p_1^4 s_2 - 2 p_1^2 s_2^2 - 4 p_1^2 s_4 \right. \nonu \\ 
& & \left. - 2 s_2 s_4 - 
     4 s_6 - \frac{2 s_2 s_6}{p_1^2} \right)  \nonu \\
& & + \frac{1}{v^4}
  \left( -5 p_1^8 + \Lambda_{N=2}^4 (3p_1^4 - 2 p_1^2 m_1^2 - 
     2p_1^2  m_2^2 +  m_1^2 m_2^2) - 8 p_1^6 s_2  \right. \nonu \\
  & & \left. \left. -3 p_1^4 s_2^2 - 6 p_1^4 s_4 - 4 p_1^2 s_2 s_4 - s_4^2 - 
     4p_1^2 s_6 - 2 s_2 s_6 + \frac{s_6^2}{p_1^4} \right ) \right] 
\eea
and the dyon vevs are given by
\bea
m_{i, dy}^4 & = & 2h^2 \left( 15 p_1^8 + \Lambda_{N=2}^4( -6p_1^4 + 3 p_1^2  m_1^2 + 
  3 p_1^2 m_2^2 -  m_1^2 m_2^2) + 20 p_1^6 s_2  \right. \nonu \\
& &  \left. + 
  6 p_1^4 s_2^2 + 12 p_1^4 s_4 + 6 p_1^2 s_2 s_4 + s_4^2 + 
  6p_1^2 s_6 + 2 s_2 s_6 \right).
\eea
Finally for meson vevs, we get
\begin{equation}
w^2_{i} = h \frac{\Lambda^{2} B}{p_1^2-m_i^2}.
\end{equation}

\subsection{$SO(2N_c+1)$ Case }

At the locus where there is one dyon which becomes massless, we have
\bea 
C_{2N_c}^2 (v^2) - \Lambda_{N=2}^{4 N_c-2- 2N_f} v^2 
\prod_{i=1}^{N_f} (v^2-m_i^2) = 
(v^2-p_1^2)^2 T(v^2).
\label{locusdyon}
\eea
The function of $w^2$ is
\bea
w^2 = \left[ H \sqrt{T(v^2)} \right]_+ \pm  H \sqrt{T(v^2)}
\eea
where $H$ is a constant.
For the finite values of $w^2$ as $t \rightarrow 0$ and $v \rightarrow \pm m_i $ we find
\bea 
w^2_i =  H \frac{C_{2N_c}(p_1^2)}{(p_1^2-m_i^2)}.
\eea
By inserting $v^2=p_1^2$ into (\ref{locusdyon}) and writing $C_{2N_c} (p_1^2) $ as
$ \Lambda_{N=2}^{2 N_c-1- N_f} p_1 
\prod_{i=1}^{N_f} (p_1^2-m_i^2)^{1/2} $, we obtain
\bea
w^2_i =  H \Lambda_{N=2}^{ 2N_c-1 - N_f} p_1 \frac{\det 
( p_1^2 - m^2)^{1/2}}{(p_1^2-m_i^2)}.
\eea
The asymptotic behavior of $w^2$ for large $v$ leads to
\bea
w^2 \sim 2 H  \frac{ C_{2N_c}(v^2)}{v^2-p_1^2} 
\sim 2 H  v^{2(N_c-1)} + 2 H p_1^2 v^{2(N_c-2)} + 
\cdots
\eea
which is the same as
$ \sum_{k=1}^{N_c} 2k \mu_{2k} v^{2(k-1)}$. From this relation we get
\bea
2H = 2N_c \mu_{2N_c} , 
\qquad p_1^2 =\frac{(N_c-1) \mu_{2(N_c-1)}}{N_c \mu_{2N_c}}.
\eea
Therefore we find $w_i^2$ completely
\bea
w^2_i =2 \sqrt{N_c ( N_c-1) \mu_{2N_c} \mu_{2(N_c-1)}} 
\Lambda_{N=2}^{2N_c-1-N_f}  
\frac{\det ( a_1^2 - m^2)^{1/2}}{(a_1^2-m_i^2)}
\eea
which exactly coincides with the meson vevs (\ref{meson1})
from field theory results in the low energy superpotential (\ref{lowsuperpo1}) in section 2.

\underline{ Example 3 : SO(5) with 1 flavor}

As we did in the case of $SO(6)$,
 the $N=2$ theory of this model  is given by the curve
\begin{equation}
t^{2} - \frac{2C_{4}(v^{2}, u_{2k})}{v} t + \Lambda_{N=2}^{4}(v^{2} - m_1^{2}) = 0
\end{equation}
where $C_{4}(v^{2}, u_{2k}) = v^{4} + s_{2}v^{2} + s_{4}$ in terms of $s_{2k}$ or
$C_{4}(v^{2}) = v^{4} - \frac{u_{2}}{2} v^{2} - \left(\frac{u_{4}}{4} -
\frac{u_{2}^{2}}{8}\right)$ in terms of $u_{2k}$.

When one dyon becomes massless, the locus in the moduli space becomes
\begin{equation}
\label{locus1}
 \left(v^{4} - \frac{u_{2}}{2} v^{2} - \left(\frac{u_{4}}{4} -
\frac{u_{2}^{2}}{8}\right)\right)^2
 - \Lambda^{4}_{N=2} v^{2} (v^{2} - m_1^{2}) = (v^{2} - p_1^{2})^2 T(v^{2}).
\end{equation}
By inserting $v^2=p_1^2$ in
equation (\ref{locus1}) and its derivative with respect to $v^2$ we obtain two equations for
$u_{2}, u_{4}$ which are solved to give
\begin{eqnarray}
\label{locus2}
u_{2} &=& 4p_1^{2} \pm \Lambda_{N=2}^{2} \left(\frac{p_1}{b} + 
\frac{b}{p_1}\right), \\ \nonumber
u_{4} &=& 4p_1^{4} \pm \Lambda_{N=2}^{2}\left(
4p_1^2\left(\frac{p_1}{b}+\frac{b}{p_1} \right)-\frac{2p_1m_1^2}{b} \right)
 + \frac{\Lambda_{N=2}^4}{2} \left(\frac{p_1}{b}+\frac{b}{p_1}\right)^2
\end{eqnarray}
where $b = \pm\sqrt{p_1^{2} - m^{2}}$ and we take only the plus sign for
$b$. 
It is easy to see that these vevs of gauge invariant variables are exactly coincident with
those of \cite{lpg,kty} obtained from the low energy effective superpotential,
for $N_c=2$,
$ \mu_{2} \mbox{Tr} (\Phi_{cl}^{2})+\mu_{4} s^{cl}_{4}  \pm 2 \Lambda_{SU(2)}^3 $ where
the classical vacua of $ \Phi_{cl}=  \sigma_2 \otimes 
\mbox{diag} (a_1, a_1, 0) $ breaks $SO(5)$ into $SU(2) \times U(1)$.
Note that the pure case has been discussed in \cite{ds} where the three 
intersection points correspond to a pair of mutually local dyons becoming massless.
We want to see the dyon vev and the finite value for $w^2$. 
The sum of two solutions of $w^2$ satisfying (\ref{solofw})
and product of them can be summarized
as
\bea
N & = & H(v^{2} + p_1^{2} -
\frac{u_{2}}{2}), \nonu \\
N^2-H^2 T & = & H^2\left(-2p_1^4+\Lambda_{N=2}^4-2s_4+p_1^2 u_2\right).
\eea
The dyon vevs are obtained from the explicit form of $T(p_1^2)$ as follows
\bea
m_{1, dy}^4= 
2 H^2 T(p_1^2)= 2H^2 \left( -2p_1b \Lambda_{N=2}^2 + \frac{m_1^4}{4p_1^2 b^2}
\Lambda_{N=2}^4 \right).
\eea
Also we have the parameter $\mu_4$
\bea
H=2 \mu_4
\eea
and the meson vevs $w^2_1$
\bea  
w^2_1= H \frac{\Lambda_{N=2}^2 p_1b}{p_1^2-m_1^2}.
\eea

\subsection { Several Massless Dyons}

We discuss the even case $SO(2N_{c})$, the odd case $SO(2N_{c} + 1)$ going 
exactly the same.
If there are $l$ massless dyons, the curve can be factorized as in (\ref{curve1})
\bea 
\label{several3}
C_{2N_c}^2 (v^2) - \Lambda_{N=2}^{4 N_c-4- 2N_f} v^4 
\prod_{i=1}^{N_f} (v^2-m_i^2) = 
\prod_{i=1}^{l} (v^2-p_i^2)^2 T(v^2)
\eea
and $w^2$ is given by:
\begin{equation}
w^2 = \left[ \frac{h(v^2)}{v^{2}}\sqrt{T(v^2)} \right]_{+} \pm \frac{h(v^2)}{v^{2}}\sqrt{T(v^2)}
\end{equation}
where now $h(v^2)$ is a polynomial of degree $2l-2$ in $v$. The degree of $h$ comes from
the degree of $w^2(v^2)$ which is $2N_{c} - 4$ 
from (\ref{boundary}) and the degree of $T(v^2)$ is $4N_{c} - 4l$
giving the degree $2l-2$ for $h(v^2)$. 
As in the case with one massless dyon, for $N_{f} < 2N_{c} - 2$ we can write 
\bea
\frac{h(v) \sqrt{T(v^2)}}{v^2} = \frac{h(v) C_{2N_c}(v^2)}{v^2 \prod_{i=1}^{l} (v^2-p_i^2)} + 
{\cal O}(v^{-4})
\eea
and we decompose
\bea 
\frac{h(v) C_{2N_c}(v^2)}{v^2} = G_{1}(p_i^{2}) +
G_{2}(v^{2}) \prod_{i=1}^{l} (v^2-p_i^2). 
\eea
We obtain $w^2_i = w^2(v^2 \rightarrow m_i^2) = \frac{G_{1}(m_{i}^{2})}{\prod_{j=1}^{l}
(m_{i}^{2} -
p_{j}^{2})}$. The value for $G_{1}(p_i^2)$ is
\begin{equation}
G_{1}(p_{i}^{2}) = h(p_{i}) \sqrt{\prod_{j=1}^{N_{f}}
(p_{i}^{2} - m_{j}^{2})} \Lambda_{N=2}^{2N_{c} - 2 - N_{f}}.
\end{equation}
Then again the discussion goes the same way as in \cite{dbo} and is similar
to the one involving only one massless dyon. We determined the coefficients
of the polynomial $G_{1}(m_{i}^{2})$
and we plugged back into the expression for $w^2_{i}$ to obtain
\begin{equation}
\label{result2}
w^2_{j} = \sum_{i=1}^{l} \frac{h(p_{i}) \mbox{det} (p_{i}^2 - m^{2})^{1/2}}
{ \prod_{k\ne i}(p_{k}^{2} - p_{i}^{2}) (p_{i}^{2} - m_{j}^{2})}
\Lambda_{N=2}^{2N_{c}-2-N_{f}}.
\end{equation}
Because the unbroken gauge group is the same as the one obtained in the 
$SU$ case, the formulas (\ref{result1}) and (\ref{result2}) agree only for
$r_{1} =\cdots = r_{l} = 2$ by identifying $a_{i}$ with $p_{i}$ and
$\phi$ with $h$. So in $SO(2N_{c})$ case as 
one of the $r_{i}$'s is greater than two, we have a disagreement between the field
theory and brane configuration result. This implies that 
$W_{\Delta} \ne 0$ in the ``integrating in" method. It would be very interesting how to
obtain the singular submanifolds of the $N=2$ Coulomb branch the low energy
effective superpotential parametrizes.

\section{Conclusions}
\setcounter{equation}{0}

In the present work we have considered $N=2$ supersymmetric gauge theories with
gauge groups $SO(N_{c})$ by using field theory approach.  
By adding a general superpotential (\ref{delw}) corresponding to
the relative orientation between two NS5 branes, we obtained the
description of the resulting $N=1$ gauge theory in both field theory and string/M theory. 
The nonsingular locus of the
$N=2$ Coulomb branch is lifted while the only singular points remain, where massless
monopoles condense by perturbation. We explicitly calculated the
monopole vevs, in field theory by studying a point in the $N=2$ moduli
space of vacua and in M theory by 
exploiting the M theory fivebrane configuration. We have obtained the same contradiction 
as in the case of $SU(N_{c})$ groups given by the low energy effective superpotential
obtained by the ``integrated in" method. In other words, this is zero for the 
enhanced gauge group $SU(2)$ but is different from zero for
$SU(r)$ with $r > 2$. We have also given the examples of $SO(5)$ and
$SO(6)$ gauge groups in order to illustrate our methods.

As in the case of the simplest mass superpotential studied previously \cite{aotsept}, 
we did not obtain any information about the particles at singular points, i.e.,
about their exact electric and magnetic charges. This remains an interesting
direction to pursue both in field theory and in M theory.
Also, it is extremely important to obtain the corrections to $W_{\Delta}$
in order to see a complete match between field theory and M theory.
It would be related to calculate the superpotential using the fivebrane configuration.
These two directions are very important and deserve further study.

\end{document}